\makeatletter
\declare@file@substitution{revtex4-1.cls}{revtex4-2.cls}
\makeatother
\documentclass[preprint]{aastex631}
\usepackage[T1]{fontenc}
\usepackage{mathtools}
\usepackage{hyperref}
\usepackage{longtable}

\shorttitle{LivRed Rotation-Age-Activity}
\shortauthors{Engle et al.}

\begin{document}

\title{Living with a Red Dwarf: X-ray, UV, and Ca \textsc{ii} Activity-Age Relationships of M Dwarfs}

\author[0000-0001-9296-3477]{Scott G. Engle}

\affiliation{Villanova University \\
Dept. of Astrophysics and Planetary Science \\
800 E. Lancaster Ave \\
Villanova, PA 19085, USA}



\begin{abstract}

The vast majority of stars in the nearby stellar neighborhood are M dwarfs. Their
low masses and luminosities result in slow rates of nuclear evolution and minimal
changes to the star's observable properties, even along astronomical timescales. 
However, they possess relatively powerful magnetic dynamos and
resulting X-ray to UV activity, compared to their bolometric luminosities. This
magnetic activity does undergo an observable decline over time, making it an important
potential age determinant for M dwarfs. Observing this activity is important for
studying the outer atmospheres of these stars, but also for comparing the behaviors
of different spectral type subsets of M dwarfs, e.g., those with partially vs.
fully convective interiors. Beyond stellar astrophysics, understanding the X-ray to
UV activity of M dwarfs over time is also important for studying the atmospheres
and habitability of any hosted exoplanets. Earth-sized exoplanets, in particular,
are more commonly found orbiting M dwarfs than any other stellar type, and thermal
escape (driven by the M dwarf X-ray to UV activity) is believed to be the dominant
atmospheric loss mechanism for these planets. Utilizing recently calibrated M dwarf
age-rotation relationships, also constructed as part of the \textit{Living with
a Red Dwarf} program \citep{2023ApJ...954L..50E}, we have analyzed the evolution of
M dwarf activity over time, in terms of coronal (X-ray), chromospheric
(Lyman-$\alpha$, and Ca \textsc{ii}), and overall X--UV (5--1700\AA) emissions.
The activity-age relationships presented here will be useful
for studying exoplanet habitability and atmospheric loss, but also for studying the
different dynamo and outer atmospheric heating mechanisms at work in M dwarfs.

\end{abstract}

\keywords{Stellar ages (1581); Stellar atmospheres (1584); Stellar coronae (305);
Stellar chromospheres (230); Stellar evolution (1599); Stellar rotation (1629);
X-ray astronomy (1810); UV astronomy (1736);
Low mass stars (2050); Late-type dwarf stars (906); M dwarf stars (982)}


\section{Introduction \& Background: Studying M Dwarfs} \label{sec:intro}

Magnetic fields are theorized to exist around all cool, main sequence stars, as massive 
as late F dwarfs, down through the late M dwarfs (collectively referred to here as
\textit{FGKM dwarfs}). These fields are responsible for (or 
contribute to) a range of observable behaviors, which include heating the outer stellar
atmosphere to 10$^5$ -- 10$^6$ K. These heated plasmas comprise the (if structured similar
to the Sun) stellar chromospheres, transition regions, and coronae. The mechanism 
responsible for generating these magnetic fields is the dynamo effect, which involves
contributions from both convective motions and stellar rotation, though in varying amounts
depending on the mass of the star (described later). The combination of FGKM dwarf
magnetic fields and stellar winds are also responsible for the spindown effect. First
discovered just over 50 years ago \citep{1972ApJ...171..565S}, this effect results in
the lengthening of an FGKM dwarf's rotation period and the weakening of its dynamo
as it ages. 

Thus, the rotation periods and/or magnetic activity levels FGKM dwarfs can
serve as age determinants. The age-rotation relationships we have constructed
for M0--6.5 dwarfs as part of the \textit{Living with a Red Dwarf} program have been
described in a companion paper \citep{2023ApJ...954L..50E}. 
This paper focuses on M dwarf activity-age relationships. 

Several observable proxies exist for measuring a star's level of magnetic activity.
Coronal emissions are regularly
characterized by observations at soft X-ray energies. The soft X-ray band encompasses
0.1 -- 10 keV although, given the sensitivity of most current X-ray instruments and
the narrow energy range over which quiescent FGKM dwarf coronae emit, most observed
fluxes are measured in a somewhat narrower energy range (e.g., 0.2--3 keV, 0.2--5 keV).
Emissions from the transition region or chromosphere
are often characterized by a range of spectral features at UV and optical wavelengths.
Even though these optical spectral features exist that probe plasmas with chromospheric 
temperatures, the majority of chromospheric emissions are observed in the UV range. For
this reason, we will use the term \textit{X--UV} to primarily refer to the outer atmospheric
(coronal--chromospheric) emissions or activity of M dwarfs. The astrophysical importance of measuring these
quantities has prompted widespread use in many studies (though this is a very active
field with far too many studies to allow a comprehensive list, a selection of recent works involving rotation rates
and/or atmospheric activity levels include \citep{2011ApJ...743...48W,2013MNRAS.431.2063S,
2016ApJ...823...16B,2016ApJ...821...81G,2017ApJ...843...31Y,2018AA...612A..89S,2020ApJ...904..140C,
2020ApJ...902....3L,2020AA...638A..20M,2021AA...649A..96J,2021ApJ...911..112Y,2023AA...671A.162M,2023AA...672A..37I}).

For this study, we have constructed X--UV activity-age relationships for M0--M6.5
dwarfs (\textit{M dwarfs} hereafter), using ages determined with the age-rotation relationships presented
in \citet{2023ApJ...954L..50E}. These relationships are not only valuable for the
information they provide about the stars themselves; they also delineate the space
environments that M dwarfs create for any exoplanets orbiting them.

This information is invaluable for multiple reasons. First is that M dwarfs represent
$\sim$75\% of the nearby stellar inventory \citep{2021AA...650A.201R} -- studying M dwarfs
allows us to understand the behavior of the universe's largest stellar population. Also,
M dwarfs are less massive ($M \approx$ 0.6 -- 0.1 $M_\odot$), smaller ($R \approx$
0.6 -- 0.1 $R_\odot$), and cooler (\textit{T}$_{\rm eff} \approx$ 3900 -- 2850 K) than the
Sun, and have much lower luminosities ($L \approx$ 0.06 – 0.001 $L_\odot$)\footnote{
\url{https://www.pas.rochester.edu/~emamajek/EEM_dwarf_UBVIJHK_colors_Teff.txt}}. Their
low masses result in slow core reaction rates and long, seemingly stable lifetimes ($\sim$100 Gyr
up to as long as $\sim$1 trillion ($\sim$10$^{12}$) years \citep{2016ApJ...823..102C}). 
While their bolometric luminosities are low, and experience little change over billions
of years, the X--UV activity of M dwarfs is comparatively strong and highly variable.
This can severely impact hosted exoplanets that would otherwise need to orbit near to the
(low luminosity) star to maintain a temperate surface. As stellar X--UV emissions are
substantial drivers of both photochemical reactions within, and loss of, planetary
atmospheres, determining the evolution of these emissions with age has achieved a 
broader impact within the field, increasing the importance of such studies.

The paper is organized as follows. In Section \ref{sec:habit}, we briefly discuss the
issues surrounding the potential habitability of planets orbiting M dwarfs. In Section
\ref{sec:rot-act}, we present the data used for this study. In Section \ref{sec:results},
the activity-age relationships will be provided and discussed. Finally, in Section
\ref{sec:conclusions}, we will provide a brief summary of our results.

\section{Orbiting M Dwarfs: Can They Host Habitable Planets?} \label{sec:habit}

Whether M dwarfs can host habitable planets is a question that has received considerable
attention from the research community. For a thorough, recent review of the numerous
factors influencing the habitability of planets orbiting G, K, and M dwarfs see \citet{2020IJAsB..19..136A}
and references therein. Although not the focus of this paper, since X--UV activity plays
a prominent role in exoplanet atmospheric evolution/loss/retention, and habitability studies, we will provide a
(very) short summary of the topic. Some of the major aspects governing potential M dwarf
exoplanet habitability can be classified as: positives, negatives, and unknowns.

`The positives' mainly involve the simple statistics. M dwarfs represent the most numerous
stellar component of observed space, live extremely long main sequence lifetimes with
near-negligible changes in either size, optical luminosity, or surface temperature, and
studies have thus far shown them to host terrestrial-size planets at a higher rate than
more massive stars. All considered, even if the odds are \textit{overall} very slim that
M dwarfs can host habitable planets, there are so many to consider that perhaps some
exoplanets have still managed to evolve with the necessary combination of characteristics
that allow them to support habitable environments. It is also worth noting that M dwarf
exoplanets are ideal targets for atmospheric composition studies via transmission
spectroscopy, due to their relatively deep transits when compare to other more massive
(luminous) spectral types. Though this last point is not a `positive' for the actual
habitability of the planet, it is a benefit when studying the planet's potential habitability
and has also generated further interest in the field.

`The negatives' center on the planetary consequences of M dwarf magnetic activity. Due to
their lower bolometric luminosities, their HZs are very close ($\sim$0.07 -- 0.4 AU for M5 -- M0 V
stars) so that tidal locking and possible tidal heating may take place \citep{2008MNRAS.391..237J}.
With the planet orbiting so near to its host star, it's exposed to more intense X--UV
emissions, along with flares and coronal mass ejections (CMEs). The end result
can be the stripping (erosion) of the planet’s atmosphere and sterilization of its surface. 

`The unknowns' primarily involve finer resolution of M dwarf activity, and at present
almost all details about the planets themselves. The more intense X--UV radiation that
potentialy habitable exoplanets around M dwarfs will receive has already been mentioned
as an overall negative. However, there is the additional unknown component of the exact role
that NUV radiation played in prebiotic chemistry on Earth \citep{2017ApJ...843..110R}. 
It is possible that M dwarfs are not \textit{luminous enough} in the NUV to drive such
chemistry, although their high flare frequencies could help compensate for this
\citep{2019ApJ...871L..26F,2023AJ....165..195B}. All
considered, much more research is still needed on these topics.
The reader is again directed to \citet{2020IJAsB..19..136A} for
a more thorough discussion, but there are numerous paths along which both the formation and
subsequent evolution of the planet and its atmosphere can proceed. M dwarfs have higher luminosities
as they are still collapsing and evolving onto the main sequence. Some studies show this
could force HZ planets into a runaway greenhouse environment, although (depending on
conditions within the nebula from which the star-planet system formed) the planets could
be protected from this. The strength/extent of the planet's magnetic field and whether it
can offer adequate protection is another important unknown \citet{2016AA...596A.111R}, as is the amount of
atmospheric replenishing that occurs due to volcanic outgassing. Related to this last
unknown are the results of \citet{2020AJ....160..237F}, proposing that planets around older and less active M
dwarfs (their study focused on GJ 699, aka Barnard's Star, age $\approx$ 9.5 Gyr and a
target within this study as well) could develop `second generation' atmospheres via
outgassing and at last become habitable worlds.

Particularly important unknowns on the stellar side are robust delineations of both M dwarf
flare frequencies and CMEs as they evolve over time. A comprehensive study by \citet{2021ApJ...915...37W} finds
that ``CMEs from M dwarfs may be much less common than generally thought, despite the high
flare rate, so perhaps CME exposure is not as big a factor for habitability as often
supposed.'' Even the habitability of a tidally locked planet can remain in question due to
unresolved planetary characteristics such as cloud cover, atmospheric circulation, and
surface arrangement.

Although the field is still some time away from precisely determining the conditions present in
a wide range of exoplanet atmospheres, surfaces, and interiors, we have at last developed a method for
determining the ages of the host stars (and thus their coeval planetary systems) and can
more precisely track the X--UV evolution of M dwarfs over time.

\section{M Dwarf Ages and X--UV Activity Measures} \label{sec:rot-act}

\subsection{A Brief Discussion of Target Ages}

We will begin by quickly summarizing the sources of M dwarf ages used in this study. The
age-rotation relationships of \citet{2023ApJ...954L..50E} were used to determine most
of the stellar ages used in this study. For a limited number of targets a reliable,
previously determined age was available and this age was used instead. These targets are
either members of star clusters/associations, belong to the Thick Disk or Halo populations
of the Milky Way, or have a companion object for which an alternative age could be determined and
applied to the M dwarf. The data initially displayed two evolutionary paths for M dwarf
rotation periods \citep{2018RNAAS...2..233G}, with the division occurring between spectral
types M2 and M2.5. This
division was confirmed as additional data were added to the relationships, and is likely
caused by the M dwarfs on either side of this divide operating
under different dynamo mechanisms. A similar
spectral type division was also found in the activity measure data of \citet{2020ApJ...891..128M}.
For ages below $\sim$2.5--3 Gyr, a further subdivision between M2.5--3.5 and M4--$\sim$6.5 
stars is apparent in the cluster rotation rate data, and this subdivision is most likely due
to the much longer pre-main sequence lifetimes of the M4--6.5 subset.

The age-rotation relationships of all M dwarf subsets show an inflection point, sharply dividing
each relationship into a first (younger) and second (older) track. This is in agreement with
previous studies (see \citealt{2020ApJ...904..140C}), who referred to the end of the younger
track as a stalling of the spindown effect, before resuming again along the older track.
A proposed theory for why this occurs is that zero age main sequence M dwarfs initially have
a strong differential rotation profile within their interiors. Along the younger track, the
stellar interior is synchronizing and transfer of angular momentum from the interior
compensates for that which is lost from the surface \citep{2020AA...636A..76S}. The younger track ends
with the re-synchronization of the stellar interior, when the star begins rotating as a solid body,
and a new track of rotational evolution begins. 

\subsubsection{A Note on Target Spectral Types}

The targets have been divided into subsets based on spectral type. As \citet{2020AA...638A..20M}
presented the largest recent database of M dwarfs with activity measures, their stellar
parameters (derived using the empirical relationships of \citet{2015ApJ...804...64M}) are used for many of the
targets in this study. However, the literature was also
searched for spectral type measures of all stars from \citeauthor{2020AA...638A..20M} as a check. 
This literature search was also used for targets presented in this study that were not
in \citeauthor{2020AA...638A..20M}. Preference was usually given to studies that either were
more recent (e.g., included updated model atmospheres), employed spectral or SED fitting, or
specifically focused on the determination of M dwarf parameters instead of a wider range of
(or all) stellar types. For all tables presented here, if (depending on the spectral type
source that was used) a target could have potentially been placed in the other M dwarf subset, 
that target's name will be in bold.

\subsection{X--UV Activity Measures}

X-rays, and most wavelengths of UV radiation, are not able to reach the Earth's surface
and thus require observations from space-based or high altitude (balloon-borne)
observatories. As these are very high quality and well-calibrated instruments, the data
they produce is second to none. Their observing time, however, is both limited and
highly sought after (oversubscribed). Though oversubscription of these instruments is
understood and unavoidable, it raises the difficulty in repeatedly measuring multiple
targets over time which would greatly benefit the X--UV relationships. The main issue is
that stellar coronal--chromospheric activity is variable over timescales of hours (flares), days to
months (stellar rotation), years (magnetic activity cycles), and finally Myr to Gyr
(weakening of the stellar magnetic dynamo due to spindown). Altogether, the amplitudes
of variation involved can be two orders of magnitude or more for the most extreme cases. Single observations of
sufficient exposure time can allow the observer to avoid contamination by stellar flares,
which can cause some of the highest-amplitude variations. The issue still remains that, until
a target has been observed multiple times over a significant time span, the observer is left
with an incomplete measure of the target’s mean activity level and this effect will
likely propagate into higher uncertainties within the activity-age relationships.

The X-ray database currently has two advantages. First, measurements don't require
high spectral (energy) resolution, allowing observations to be gathered more efficiently.
Second, the German--US--UK collaborative R\"ontgen Satellite (R\"OSAT) was designed for
efficient X-ray observing and carried out numerous observations of stars in the 1990s,
including an invaluable all-sky survey. Observations with modern X-ray missions
(e.g. \textit{Chandra}, \textit{XMM-Newton}, \textit{Neil Gehrels Swift}, and very recently
\textit{eROSITA}) can be added to this previous data if available and begin to better characterize the range
of activity (and mean activity level) for each target.

The majority of UV activity from cool dwarfs can be attributed to emission lines. Specifically for M
dwarfs, a single feature (the Lyman-$\alpha$ 1216\AA~ [Ly$\alpha$] line) is responsible for 50\%
of the total UV emissions or more \citep{2012ApJ...750L..32F}. Due to its relative strength, this line has
naturally become a high-priority target for observers, but a successful analysis
requires data with medium to high spectral resolution and/or an instrument that can
successfully mitigate geocoronal contamination. Interstellar absorption
features of hydrogen and deuterium occur at the same wavelengths, and these need to
be modelled and accounted for before the `pure' stellar Ly$\alpha$
flux can be obtained. The instrument and analysis requirements result in a more limited
database of reliable Ly$\alpha$ measures, and most of those presented here were obtained
from the literature and measured in the previously described fashion by the \textit{MUSCLES}
and \textit{mega-MUSCLES} \textit{HST} observing programs \citep[see][]{2016ApJ...820...89F,2016ApJ...824..101Y,2016ApJ...824..102L,2019ApJ...871L..26F,2021ApJ...911...18W}.

Fortunately, there are also a number of cool dwarf magnetic activity `tracers' observable at optical
wavelengths. As in the UV, this activity is observed through emission (sometimes absorption) lines. 
Though popular alternatives exist (e.g., H$\alpha$ and Na \textsc{i}), we have selected the
Ca \textsc{ii} H \& K features for this study. The behavior of these features in cool stars
has been studied for over 40 years \citep{1980PASP...92..385V}, and they occur at the short
end of the optical spectrum ($\sim$3930--3970\AA), where M dwarf photospheric contributions
will be minimal. Even so, a reliable and well-studied quantity ($R'_{HK}$) has been derived for these features
which involves removing the photospheric contribution to allow an easier comparison of different
spectral types. Still, a recent study by \cite{2023AA...671A.162M} found that offsets can occur
depending on the exact methodology used over time. When applicable, the corrections of
\citeauthor{2023AA...671A.162M} were applied to target data before the representative value
presented here was determined. After applying the correction, comparing the values determined in
this study to those of \citeauthor{2023AA...671A.162M}returned an average difference of
$\Delta\log R'_{HK} \approx$ 0.031.

The clear advantage of optical magnetic activity tracers like Ca \textsc{ii} HK
is that they can be observed using ground-based instruments with lower oversubscription rates. Still,
medium- to high-resolution spectroscopy is required and the faintness of the targets can be
prohibitive for smaller telescopes, especially when observing M dwarfs at short optical
wavelengths. Previous spectroscopic surveys were dedicated to such activity measures, but recent
planet-hunting surveys (e.g., with the \textit{HARPS} instrument) have produced large 
numbers of high quality spectra that have also been used for activity measures, many of which
were included in this study (see \citealt{2021AA...652A.116P}).

\section{The M Dwarf Activity-Age Relationships} \label{sec:results}

\begin{figure}[ht!]
\plotone{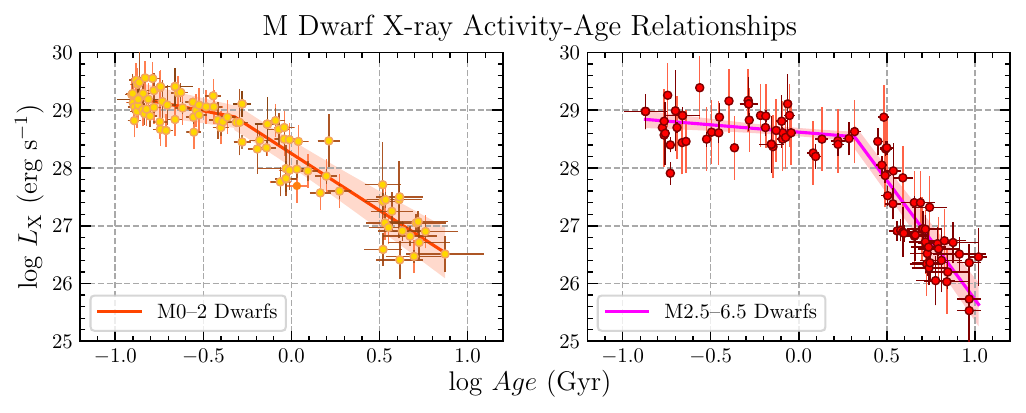}
\caption{The evolution of M dwarf X-ray activity over time. Early M dwarfs are plotted on the left, and mid-late M dwarfs
on the right. Lighter-colored error bars indicate when the `average' uncertainty value was applied to a target for fitting
purposes. A two-component, segmented linear model was applied to each subset and is plotted. Each subset shows an initial
``saturation phase'' where high activity levels are sustained for a period of time, before an inflection point is reached.
After this, the activity decreases at an accelerated rate. It is worth noting that many studies assume a constant level of
activity during the saturation phase, where our fits indicate that a slight decrease occurs. The saturation phase is found
last $\sim$470 Myr for the early M dwarfs and $\sim$2.1 Gyr for the mid-late M dwarfs. \label{fig:lxage}}
\end{figure}

\begin{figure}[ht!]
\plotone{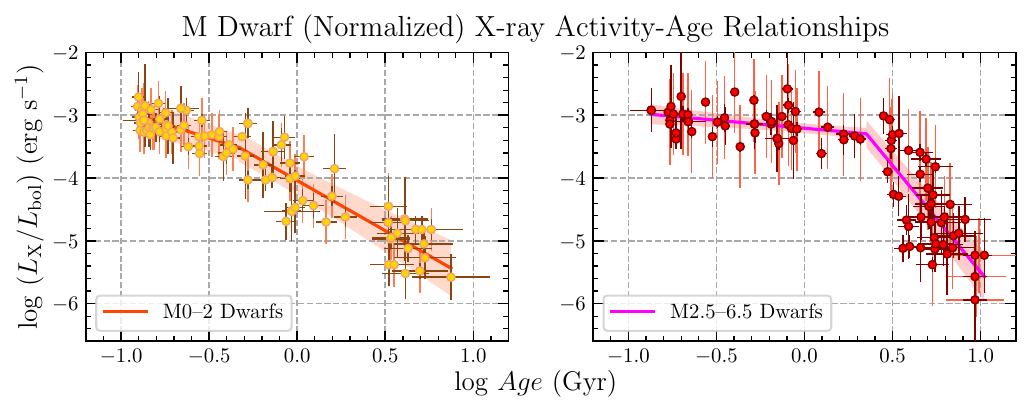}
\caption{The evolution of M dwarf X-ray activity over time is again plotted, but now the ratio of X-ray to bolometric
luminosity is used to better normalize the stars. Again, lighter-colored error bars indicate when the `average'
uncertainty value was applied to a target for fitting purposes. As revealed by the segmented fits to the data, both subsets begin
their saturation phases at essentially equal activity levels ($\log (L_{\rm{X}}/L_{\rm{bol}}) \approx -3$), and decline
by $\sim$2.5 orders of magnitude. The length of the saturation phase is found to be $\sim$450 Myr for the early M dwarfs,
but $\sim$2.3 Gyr for the mid-late M dwarfs.\label{fig:lxbolage}}
\end{figure}

\begin{figure}[ht!]
\plotone{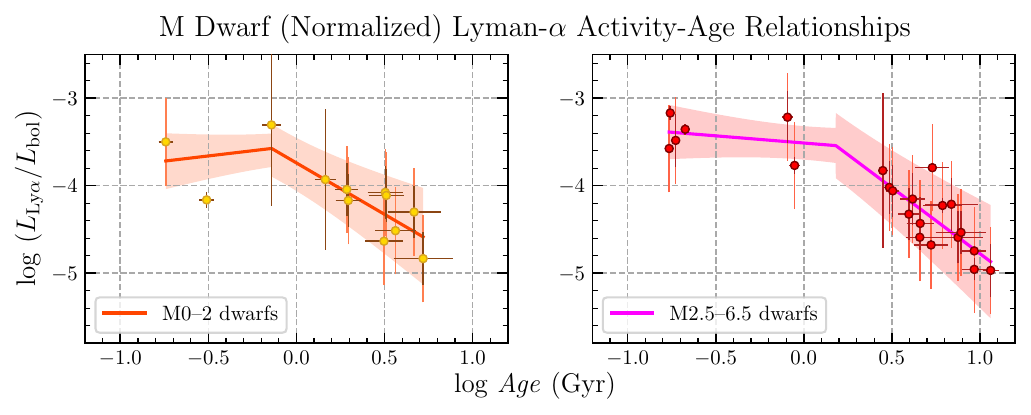}
\caption{The evolution of Lyman-$\alpha$ activity over time is plotted. (Again, lighter-colored error bars indicate when
the `average' uncertainty value was applied to a target for fitting purposes.) The reliability of the fits is impacted by the
lack of data, particularly for the early M dwarfs. The current data show saturation phase lengths of $\sim$720 Myr for
early M dwarfs and $\sim$1.5 Gyr for mid-late M dwarfs. Again, however, these quantities should not be regarded as highly
reliable due to the lack of data. Instead, we would advise using the Ca \textsc{ii} relationship (Fig.\ref{fig:caiiage})
as a more reliable indicator of chromospheric activity over time. \label{fig:lyaage}}
\end{figure}

\begin{figure}[ht!]
\plotone{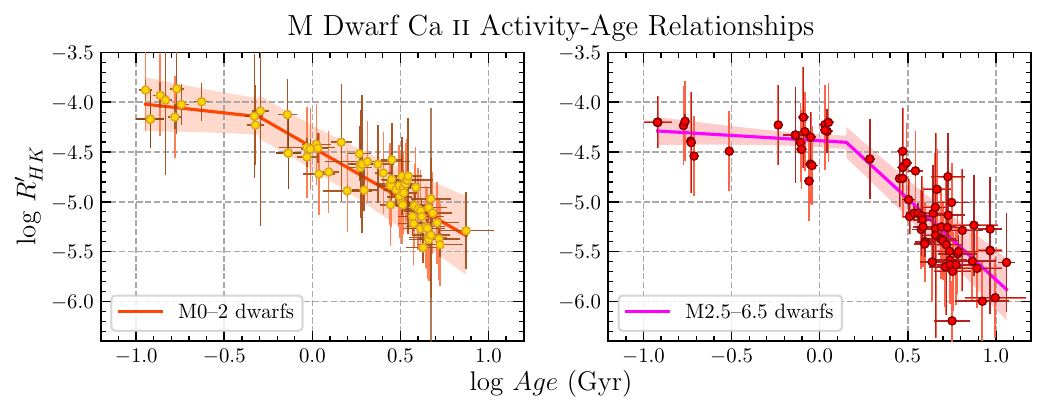}
\caption{Evolution of Ca \textsc{ii} $HK$ emission over time, as given by the $\log R'_{HK}$ index. (Once again,
lighter-colored error bars indicate when the `average' uncertainty value was applied to a target for fitting
purposes.) The Ca \textsc{ii}
emission line cores form within the stellar chromosphere and can thus serve as a proxy for cooler UV features, such as
Lyman-$\alpha$. Given the richer dataset, the Ca \textsc{ii} relationships are also more reliable than those for
Lyman-$\alpha$. Both stellar subsets are observed to decline by $\sim$1.5 orders of magnitude, a notably less drastic
decline than shown for X-ray activity, with saturation phase lengths of $\sim$500 Myr for early M dwarfs and $\sim$1.4
Gyr for mid-late M dwarfs. The early M dwarf saturation phase length can be considered comparable to the X-ray, given
the uncertainties. The mid-late M dwarfs show a considerable difference in saturation phase lengths \textit{at present}.
It's important to note, however, that although the Ca \textsc{ii} dataset is far richer than Lyman-$\alpha$, one
age-range that is not well-covered occurs right near the inflection point. Further data at this vital age-range will
be necesary to reliably pin down the chromospheric saturation phase length in comparison to the coronal, as shown by
the X-ray data. Another notable feature of these relationships is that early M dwarfs appear to exhibit higher relative
levels of chromospheric activity at both young and old ages. However, much of this difference can be accounted for by
the uncertainties and scatter of the fits and data. \label{fig:caiiage}}
\end{figure}

Here we present the relationships constructed by the \textit{Living with a Red Dwarf}
(\textit{LivRed}) program, along with a discussion of the methods used to develop them, and their
results. As was done previously for the rotation-age relationships, a two-segment linear
equation was defined via \texttt{numpy.piecewise} and then fit to each data set using
\texttt{scipy.optimize.least\_squares}. Within the sample of each activity index, an average
`scatter' value was calculated using the targets with multiple measures, and this value was
used as the uncertainty for single-measure targets when fitting the data. This application of
a larger uncertainty value than most targets' data would indicate assumes that all targets
have variable activity levels, but only those with multiple measures have given an insight
into what the average breadth of those variations should be. In Figs
\ref{fig:lxage}, \ref{fig:lxbolage}, \ref{fig:lyaage}, \ref{fig:caiiage}, the lighter error bars show instances
where this average scatter value has been applied, and the darker error bars show the
uncertainty based on actual data of the target itself.

Activity-Age relationships have been constructed in the X-ray, UV, and for the short optical
wavelength Ca \textsc{ii} emission line doublet. The Ca \textsc{ii} observational database is much
richer than that of Ly$\alpha$, and the two emission features form in plasmas of overlapping
temperatures and stellar atmospheric regions (though the Ly$\alpha$ region extends to higher
temperatures/altitudes, while the Ca \textsc{ii} regions extends to lower temperatures/regions
-- \citealt{1981ApJS...45..635V,2008LRSP....5....2H}). These relationships serve two purposes:
they provide additional age determinants for M dwarfs, as we will discuss, but they also delineate
the average high energy (X--UV) fluxes that exoplanets orbiting such stars have been, are
presently being, or will in the future be, subjected to. Further, thoroughly studied relationships
exist between Ca \textsc{ii}, Ly$\alpha$, and numerous other UV 
emission features (e.g., see \citealt{2020AJ....160..269M,2021ApJ...911..111P} and references therein). 
These could allow any or all of the relationships (or combinations thereof) presented both here and in
\citet{2023ApJ...954L..50E} to be utilized in refining an M dwarf's age,
and the high energy environment that an exoplanet orbiting it would be subjected to.

The X-ray relationships are plotted in Fig \ref{fig:lxage} (X-ray luminosity over
time) and Fig \ref{fig:lxbolage} (the ratio of X-ray to bolometric luminosity over time), and the
best-fitting parameters are presented in Table \ref{table:parameters}.
We have endeavored to gather multiple measures of X-ray activity for
each star whenever possible, to help mitigate scatter due to stellar variability. This does not
eliminate the issue as many of the stars still only possess 1 or 2
measures. As further measures of the targets are carried out by current and future X-ray missions, it
will be interesting to see how (or if) the remaining scatter in Figs \ref{fig:lxage} and
\ref{fig:lxbolage} is reduced. M dwarf ages were calculated using the Age-Rotation relationships
presented in \citet{2023ApJ...954L..50E}, unless there was an existing age determination
(e.g., membership in a cluster or the Thick Disk/Halo populations).
As seen in the plots, the sample is separated into the same `early' and `mid-late' subsets as
for the rotation relationships, and the X-ray
activity of both M dwarf subsets decreases by $\sim$2.5--3 orders of magnitude over the plotted age
range ($\sim$0.1 -- 12.5 Gyr). 

The most notable difference between the M dwarf subsets is the substantially shorter `saturation
phase' of early M dwarfs, compared to the mid-late subset. Though many previous studies fixed
the saturation phase to a flat level of activity, we instead opted to leave all parameters as free.
In both subsets, there appears to be a slight decline in activity over the duration of the saturation
phase, though it is within the scatter of the data. As with the age-rotation relationships, we
advise restraint when using the first `track' of the X-ray relationships to quote a precise age.
Here, however, is where one aspect of the early subset relationship's usefulness is found. After
$\sim$0.5 Gyr, early M dwarfs exit their saturation phase and X-ray activity levels begin to decline
more apidly. This establishes a range of ages where, although the \textit{rotations} of early M
dwarfs are still evolving along the less reliable, first (re-synchronizing) track, their X-ray
activity has shifted onto its second track, where more reliable ages can be determined. In effect, the
X-ray relationship extends the range over which more reliable ages can be determined for early M dwarfs.
Additional methods of age determination involving kinematics \citep{2021AJ....161..189L} and 
abundance \citep{2023ApJ...942...35C} can also be employed to achieve results based on as many
techniques as the data allows. This opportunity to determine M dwarf ages through multiple methods,
or to compare the ages of each method for targets with high-quality data, holds great potential
for future studies.

Although the mid-late subset appears to begin its saturation phase at a slightly lower X-ray
luminosity than the early subset ($\sim$28.9 vs. 29.2 erg s$^{-1}$ cm$^{-2}$, according to the
fits [\ref{fig:lxage}], though within the scatter), it's important to note that both subsets
show almost the same initial levels of X-ray activity when normalized according to bolometric
luminosity (see Fig. \ref{fig:lxbolage}). Also, the mid-late M dwarfs remain in their saturation
phase for as long as $\sim$2.3 Gyr. This presents a far more extended period of time over
which mid-late M dwarfs sustain enhanced levels of X-ray activity, making the ages determined
via this method less reliable over a larger age-range, and any exoplanet orbiting a mid-late
M dwarf must endure an initial phase of high X-ray activity that is $\sim4-5\times$ longer than
it would be if the planet were orbiting an early M dwarf. The consequences of this phase on the
planet's atmosphere, even potentially affecting its ability to retain elements heavier than
H/He (see \citealt{2016A&A...596A.111R,2016PhR...663....1S}), would be particularly relevant to
M dwarfs and stands to benefit from additional theoretical work.

This dramatic difference in saturation phase length is also interesting in terms of stellar astrophysics. Early
M dwarf X-ray activity begins to decline shortly after they arrive on the main sequence, but mid-late M dwarf
X-ray activity remains saturated until they are nearly finished evolving along their first rotation track.
This is likely a result of the different dynamo mechanisms at work within the two subsets, and the relative
contributions of rotation/convection towards each one. As described in \citet{2020ApJ...891..128M}, there
are three stellar magnetic dynamos
for which models exist -- $\alpha\Omega$, $\alpha^2$, and $\alpha^2\Omega$ (where $\alpha$ represents the contribution 
due to convective turbulence (in the presence of rotation, giving rise to Coriolis force effects), and $\Omega$ represents angular velocity). The more
massive early M dwarfs are theorized to operate under the $\alpha\Omega$ dynamo -- similar to the Sun -- and
this would explain why the spindown effect, and decreasing angular velocity, would quickly result in weakening
magnetic fields and lower X-ray activity levels. The mid-late M dwarfs, however, generate magnetic fields via
either the $\alpha^2\Omega$ or $\alpha^2$ dynamo. Due to the diminished contribution of angular
velocity, mid-late M dwarfs can evolve though $>$ 2 Gyr of spindown, and associated angular velocity loss, 
without any substantial weakening of their magnetic fields. Eventually, though, rotation slows to the point where
it sufficiently impacts the convective turbulence ``$\alpha$ effect'' mechanism, the 
magnetic field begins to weaken, and X-ray activity begins to decline. However, this is simply a qualitative
description of how the dynamo models may explain the data and relationships as presented. A full, theoretical
explanation is outside the scope of the current paper.

\begin{deluxetable*}{ccc|cc}
\tablecaption{Best-fitting Activity-Age parameters. \label{table:parameters}}
\tablehead{\colhead{}  & \multicolumn{2}{c}{M0--2}
           & \multicolumn{2}{c}{M2.5--6.5}\\
           \colhead{} & \colhead{parameter} & \colhead{uncertainty}  
           & \colhead{parameter} & \colhead{uncertainty}}
\startdata
  & \multicolumn{4}{c}{$\log L_{\rm X}$}          \\
\hline
a & -0.5836 & 0.2447 & -0.2534 & 0.1505 \\
b & 28.6993 & 0.1787 & 28.6213 & 0.0726 \\
c & -1.3858 & 0.2689 & -3.8713 & 0.3988 \\
d & -0.3240 & 0.0902 & 0.3171  & 0.0487 \\
\hline
  & \multicolumn{4}{c}{$\log L_{\rm X}/L_{\rm bol}$}     \\
\hline
a & -0.8820 & 0.2444 & -0.2542 & 0.1591 \\
b & -3.8021 & 0.1784 & -3.2064 & 0.0768 \\
c & -0.7164 & 0.2685 & -3.1416 & 0.4184 \\
d & -0.3314 & 0.1729 & 0.3516  & 0.0611 \\
\hline
  & \multicolumn{4}{c}{$\log L_{\rm Ly\alpha}/L_{\rm bol}$}   \\
\hline
a & 0.2081  & 1.7016 & -0.1619 & 0.3197 \\
b & -3.5663 & 0.3692 & -3.5114 & 0.1914 \\
c & -1.3478 & 0.9079 & -1.3472 & 0.4673 \\
d & -0.1420 & 0.3436 &  0.1809 & 0.2333 \\
\hline
  & \multicolumn{4}{c}{$\log R'_{HK}$}        \\
\hline
a & -0.1929 & 0.2205 & -0.1062 & 0.1354 \\
b & -4.2022 & 0.1713 & -4.3860 & 0.0648 \\
c & -0.8214 & 0.2329 & -1.5184 & 0.2502 \\
d & -0.3053 & 0.1487 &  0.1512 & 0.0942 \\
\hline
  & \multicolumn{4}{c}{$\log (L_{X-UV (5-1700\mathring{\mathrm{A}})} / L_\textrm{bol})$}        \\
\hline
a & -0.4896 & 0.1266 & -0.1456 & 0.0911 \\
b & -3.2128 & 0.0914 & -2.8876 & 0.0439 \\
c & -0.4469 & 0.1441 & -1.8187 & 0.2412 \\
d & -0.2985 & 0.1528 & 0.3545  & 0.0604 \\
\hline
\enddata
\tablecomments{\footnotesize{All equations are of the format: \\
Activity index = $a \times \log Age (Gyr) + b~~~[for \log Age < d]$ \\
Activity index = $a \times \log Age (Gyr) + b + c\times(\log Age(Gyr) - d)~~~[for \log Age \geq d]$
}}
\end{deluxetable*}

For a direct measure of UV activity, the Ly$\alpha$ emission line has been used, as it represents
the majority of all UV flux emitted from cool stars, especially so for M dwarfs.
The Ly$\alpha$ database consists of far fewer measures compared to the X-ray, though it is no less
important as Ly$\alpha$ probes a different layer of the stellar atmosphere and allows accurate stellar
wind measures. Obtaining these are crucial for gauging the likelihood that an exoplanet will
be able to retain its atmosphere. Fig \ref{fig:lyaage} plots the evolution of
Ly$\alpha$ activity over time, using the same early and mid-late M dwarf subsets as for the
X-ray relationships. As with the X-ray relationships, the best-fitting parameters for the 
Ly$\alpha$ data are provided in Table \ref{table:parameters}. The lack of data is immediately
visible in the plots, and the increased
uncertainty of the fits. The inflection points are more poorly defined, as is the saturation
phase for the early subset where only 3 measures define this first track. Consequently, this
diminishes the reliability of the fit for the second track, as well, though the current fit for this
phase is not unreasonable for such a limited dataset. For the mid-late 
M dwarfs, a similar lack of data near the transition between the saturation and declining-activity phases 
leaves ambiguity in the determination of the inflection point. However, the fit does appear to track the
data for the older stars, which makes the current estimate of $\sim$1.5 Gyr for the duration of the
saturation phase a reasonable one. With the limited data currently available, the mid-late subset
appears to have a slightly higher level of Ly$\alpha$ activity, relative to bolometric luminosity, when
compared to the early subset. This agrees with \citet{2020ApJ...902....3L}, who also reported a
trend of increasing activity with decreasing effective temperature.



The final activity index we have analyzed for this study is the $\log R'_{HK}$ index, derived
from the Ca \textsc{ii} H \& K emission features at $\sim$3968.5\AA~ and $\sim$3933.7\AA,
respectively. As this measure can be obtained from the ground, there is a greater availability
of instruments and observing time. As a result, the Ca \textsc{ii} database is much more
extensive than Ly$\alpha$, and begins to rival the X-ray database, even though there haven't yet
been any sensitivity-limited all-sky Ca \textsc{ii} surveys for M dwarfs as R\"OSAT carried out
in the X-ray regime. Due to the large-scale spectroscopic surveys searching for planets around
low-mass stars that have come online in recent years, a particular advantage of the Ca \textsc{ii}
database is that it contains a substantial sample of targets with time-series measures. Many of these 
surveys have determined or confirmed rotation periods for their M dwarf targets using the repeat
measures of activity indices such as Ca \textsc{ii}. The evolution of M dwarf Ca \textsc{ii} activity
over time is plotted in Fig. \ref{fig:caiiage} (again, using the same M dwarf subsets as for the
X-ray and Ly$\alpha$ relationships) and the best-fitting paramaeters are provided in Table \ref{table:parameters}.



As Fig. \ref{fig:caiiage} shows, the richer data sets allow for a better understanding of the
levels of activity during the saturation phase. According to the fits, early M dwarfs appear
slightly more active during both their saturation phase and at very old ages. Both subsets
show a $\sim$1.5 order of magnitude decrease in Ca \textsc{ii} activity; less than
the decrease they experience in X-ray activity. This phenomenon of chromospheric activity experiencing
a less drastic decline with age has also been observed in other studies 
\citep{2020ApJ...902....3L,2021ApJ...907...91L,2021ApJ...911..111P}.
One potential theory put forth is that cool star chromospheres are acoustically heated, and the
coronae are magnetically heated. This hints at a potentially interesting scenario where the
chromospheres and coronae of early and mid-late M dwarfs are, when compared to each other, declining
by the same relative amounts even though \textit{a)} one subset is partially convective and the other is
fully convective, and \textit{b} the chromospheres are acoustically heated (driven by convection), yet the
coronae of the two subsets should be heated by different convection/rotation contributions. In studying
several FUV emission features, \citet{2021ApJ...907...91L} further found that transition region
activity levels may behave intermediate to those of the chromosphere and corona, perhaps indicating
how the different heating contributions vary between the atmospheric layers. \citet{2020ApJ...902....3L}
found that FGKM stars all display similar levels of saturated coronal activity, but as they age they
display an intensifying trend of inceasing coronal activity with decreasing effective temperature.
As mentioned previously, \citeauthor{2020ApJ...902....3L} also reported a trend of increasing Ly$\alpha$
activity with decreasing effective temperature, but the trend was essentially consistent across stellar
age and not as steep as the coronal activity trend for their oldest target group. Further,
\citeauthor{2020ApJ...902....3L} proposed additional theories, including that flare heating could
play a more prominent role in the atmospheres of lower mass stars, or that the higher surface
gravity of M dwarfs could lead to different atmospheric structures, higher photospheric gas
pressure and, in turn, stronger magnetic fields.

The chromospheric activity
of early M dwarfs exits the saturation phase after $\sim$600 Myr, where coronal activity exits after $\sim$500 Myr. 
This difference is not conclusively determined and the chromospheric and coronal activity levels
can be considered to exit the saturation phase at a similar age to each other.
The mid-late M dwarf inflection point is not strictly defined by the fits,
since the Ca \textsc{ii} database also suffers from a (albeit less substantial) lack of data near this
age-range. For this subset, the Ca \textsc{ii} data show that chromospheric activity levels exit
the saturation phase at $\sim$1.4 Gyr, where coronal activity levels take $\sim$2.3 Gyr to exit this
phase. This is a more substantial difference, but again one that is owed heavily to the
lack of near-inflection Ca \textsc{ii} data. Further data at this important age-range could
reduce, possibly even eliminate, the current disparity between the lengths of the chromospheric
and coronal saturation phases. Firmly establishing the chromospheric and coronal inflection points
can help shed light on the outer atmospheric heating mechanisms at work in M dwarfs. It is encouraging
to see that the chromospheric saturation phase lengths derived from Ca \textsc{ii} data are close to
those obtained from the much more sparse Ly$\alpha$ data. 

To better aid planetary atmosphere and habitability studies, we have converted our X-ray
($\log (L_X / L_{bol})$) relationships into cumulative X-ray to Ultraviolet (X--UV: $\sim$5 -- 1700\AA) irradiance relationships. To do this, spectral energy distributions (SEDs) constructed
by the \textit{MUSCLES} and \textit{Mega-MUSCLES} surveys (\citealt{2016ApJ...820...89F,2016ApJ...824..101Y,2016ApJ...824..102L,2019ApJ...871L..26F,2021ApJ...911...18W}) were obtained from MAST\footnote{\url{https://archive.stsci.edu/prepds/muscles/}}
\citep{https://doi.org/10.17909/t9dg6f} and
used to determine integrated fluxes over the 5 -- 1700\AA~ range. A linear relationship was determined
to be:

\begin{equation}
    \log (L_{X-UV (5-1700\mathring{\mathrm{A}})} / L_\textrm{bol}) = 0.5728 [0.0589] \times \log (L_{\mathrm{X}} / L_\textrm{bol}) - 1.0509 [0.2921]
\end{equation}

Using this equation, we were able to analyze the X--UV irradiances of M dwarfs over time and 
determine the segmented fit parameters shown in Table \ref{table:parameters}. These
relationships have not been plotted, as they mirror those of the $\log (L_X / L_{bol})$ activity over time
(save for the relative shifting/scaling of values). At present, since X-ray data were used as the
basis for scaling, the saturation phases again last for $\sim$500 Myr and $\sim$2.3 Gyr for the early
and mid-late M dwarfs, respectively.

\section{Conclusions} \label{sec:conclusions}

\begin{figure}[ht!]
\plotone{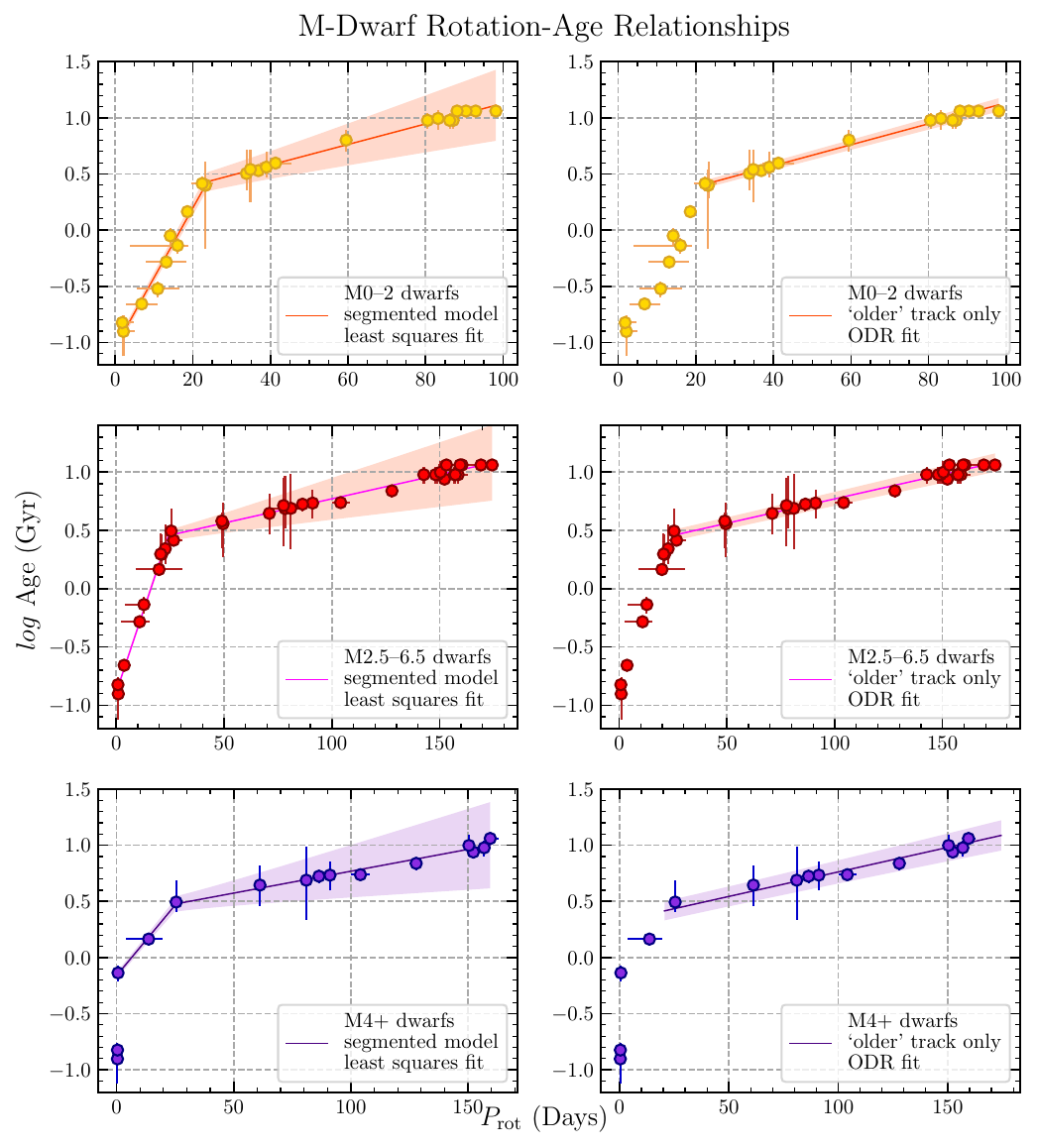}
\caption{M dwarf age-rotation relationships from the \textit{Living with a Red Dwarf} program \citep{2023ApJ...954L..50E}.
With the exception of cluster or galactic population members, these relationships were used to calculate the stellar ages
used in this study and presented in the Tables. Each row of plots is devoted to a specific subset of M dwarfs. The `early'
M dwarfs, M0--2 dwarfs, are plotted in the top row. The `mid-late' M dwarfs are plotted in the middle row. For the `older'
track, M2.5--6.5 dwarfs are plotted together since they have all settled onto a common evolutionary path. However, only
M2.5--3.5 dwarfs are plotted on the `young' track due to the large differences in pre-main sequence lifetimes that are
encountered within this mass-range. Consequently, a third subset was created and shown in the bottom row. Here, only
M4 (and later) dwarfs are plotted. As seen, there is a large difference between the young tracks of the middle and bottom
rows, but any difference between the older tracks is well within the uncertainties of the fits. \label{fig:rotagesemilog}}
\end{figure}

The \textit{Living with a Red Dwarf} (\textit{LivRed}) program has constructed activity-age relationships for M
dwarfs using three different tracers of magnetic activity, along with a broad X-ray to Ultraviolet
band (X-UV: 5--1700\AA). Multiple measures of each star's activity levels were gathered whenever
possible to help mitigate scatter due to stellar variability, and many of the ages were calculated
using the age-rotation relationships presented in \citet{2023ApJ...954L..50E} and shown in Fig.
\ref{fig:rotagesemilog}.

The stars were divided into two subsets, as previously observed 
\citep{2018RNAAS...2..233G,2020ApJ...891..128M} and utilized when constructing the age-rotation
relationships -- what we have called the ``early'' [M0--2] and ``mid-late'' [M2.5--6.5] subsets.
Both subsets show similar drops in relative activity levels as they evolve.
The most significant difference between the two subsets' activity levels is the
duration of their initial saturation phases. Early M dwarfs experience a coronal (X-ray) and 
chromospheric (Ca \textsc{ii}) saturation phase of $\sim$500--600 Myr. Mid-Late dwarfs, by
contrast, experience a coronal saturation phase of $\sim$2.3 Gyr and a chromospheric saturation
phase that is currently estimated to last $\sim$ 1.4--1.5 Gyr, but is not as well constrained by
either Ca \textsc{ii} or Ly$\alpha$ data. As proposed by other studies in the literature, the
disparity in coronal vs. chromospheric activity behaviors for mid-late M dwarfs could
\textit{potentially} indicate that the atmospheric
plasmas are being heated by different mechanisms, and further study along these lines is
encouraged.

For both subsets, however, the extended initial periods of enhanced X--UV activity will be
important to account for when 
analyzing the potential habitability of any planets orbiting M dwarfs. For example, the publicly
available \texttt{VPlanet}\footnote{\url{https://github.com/VirtualPlanetaryLaboratory/vplanet}} software
has a number of useful routines dealing with exoplanet habitability calculations
\citep{2022ApJ...928...12D}, but the default prescription for high-energy stellar evolution is
that of \citet{2005ApJ...622..680R} which was constructed for solar-type G dwarfs and uses
a saturation phase length of 100 Myr. This can significantly underestimate the saturation phase
of M dwarfs, and the data and relationships presented here will help future studies to better
account for such differences.

\begin{longrotatetable}
\begin{deluxetable*}{lcccccccccc}
\tablecaption{Early M Dwarf X-ray Data\label{table:earlyxraydata}}
\tablehead{\colhead{Star Name}  & \colhead{$P_{\rm{rot}}$}  & \colhead{err}
           & \colhead{$P_{\rm{rot}}$ src}  & \colhead{$\log$ Age}  & \colhead{err}  
           & \colhead{$\log(\langle L_{\rm X}\rangle$)}  
           & \colhead{err}  
           & \colhead{$\log(\langle L_{\rm X}\rangle$ / $L_{\rm bol})$}
           & \colhead{err}  & \colhead{X-ray src}}
\startdata
1RXS   J041330.7+251052 & 2.22  & 0.02 & M20  & -0.906 & 0.038 & 29.27 & 0.08 & -2.86 & 0.10 & M20        \\
\textbf{EPIC 201909533}          & 2.29  & 0.02 & M20  & -0.901 & 0.038 & 29.29 & 0.33 & -2.71 & 0.40 & R, M22     \\
1RXS J030708.3+323528   & 2.35  & 0.02 & M20  & -0.898 & 0.038 & 29.13 & 0.06 & -3.03 & 0.07 & M20        \\
\textbf{EPIC 210651981}          & 2.44  & 0.02 & M20  & -0.892 & 0.038 & 28.82 & 0.05 & -3.24 & 0.06 & M20        \\
1RXS J231233.3+391140   & 2.47  & 0.02 & M20  & -0.890 & 0.038 & 29.06 & 0.11 & -3.12 & 0.13 & M20        \\
\textbf{GSC 03517-00212}         & 2.61  & 0.03 & M20  & -0.882 & 0.039 & 29.52 & 0.02 & -2.92 & 0.02 & M20        \\
\textbf{UCAC4 630-046962}        & 2.69  & 0.03 & M20  & -0.877 & 0.039 & 28.98 & 0.03 & -2.98 & 0.04 & M20        \\
CK Tri                  & 2.73  & 0.03 & M20  & -0.874 & 0.039 & 29.45 & 0.01 & -2.84 & 0.01 & M20        \\
LP 152-39               & 2.79  & 0.03 & M20  & -0.870 & 0.039 & 29.09 & 0.03 & -3.27 & 0.04 & M20        \\
Pleiades/NGC 2516       &       &      &      & -0.870 & 0.120 & 29.19 & 0.28 & -3.08 & 0.27 & N16        \\
DT Vir (GJ 494)         & 2.89  & 0.03 & 1T   & -0.864 & 0.039 & 29.48 & 0.55 & -2.85 & 0.66 & R, X       \\
UCAC4 587-034710        & 3.05  & 0.03 & M20  & -0.854 & 0.039 & 29.24 & 0.09 & -3.24 & 0.11 & M22        \\
UCAC4 681-049380        & 3.22  & 0.03 & M20  & -0.844 & 0.039 & 29.01 & 0.04 & -3.12 & 0.05 & M20        \\
1RXS J085922.2+403351   & 3.25  & 0.03 & M20  & -0.842 & 0.039 & 29.29 & 0.25 & -3.20 & 0.30 & M20        \\
OT Ser (GJ 9520)        & 3.37  & 0.03 & M20  & -0.834 & 0.039 & 28.94 & 0.19 & -3.31 & 0.24 & R, X       \\
HAT 169-07364           & 3.42  & 0.03 & M20  & -0.831 & 0.039 & 29.56 & 0.07 & -2.91 & 0.08 & M20        \\
PM J23045+4014          & 3.56  & 0.04 & M20  & -0.823 & 0.039 & 29.02 & 0.05 & -3.20 & 0.06 & M20        \\
EPIC 206349327          & 3.72  & 0.04 & M20  & -0.813 & 0.039 & 29.19 & 0.04 & -3.21 & 0.05 & M20        \\
\textbf{G 80-21}                 & 3.88  & 0.04 & M20  & -0.803 & 0.039 & 28.90 & 0.19 & -3.19 & 0.23 & M22        \\
G 173-18                & 4.12  & 0.04 & M20  & -0.788 & 0.039 & 29.55 & 0.01 & -2.81 & 0.01 & M20        \\
FK Aqr (GJ 867 A)       & 4.23  & 0.04 & M20  & -0.781 & 0.039 & 29.34 & 0.30 & -3.06 & 0.36 & M20        \\
2MASS J05071824+5307115 & 4.23  & 0.04 & M20  & -0.781 & 0.039 & 29.05 & 0.09 & -3.26 & 0.11 & M20        \\
HAT 222-10112           & 4.77  & 0.05 & M20  & -0.747 & 0.040 & 29.39 & 0.06 & -2.96 & 0.07 & M20        \\
2MASS J04160235+3325301 & 4.77  & 0.05 & M20  & -0.747 & 0.040 & 28.80 & 0.05 & -3.22 & 0.06 & M20        \\
BD-17 6768              & 4.83  & 0.05 & M20  & -0.744 & 0.040 & 28.67 & 0.03 & -3.33 & 0.04 & M20        \\
AU Mic (GJ 803)         & 4.85  & 0.05 & M20  & -0.743 & 0.040 & 29.41 & 0.07 & -3.15 & 0.12 & C, R, S, X \\
EPIC 202059229          & 5.01  & 0.05 & M20  & -0.733 & 0.040 & 29.15 & 0.37 & -3.17 & 0.44 & M22        \\
\textbf{GSC 02083-01558}         & 5.34  & 0.05 & M20  & -0.712 & 0.040 & 28.65 & 0.06 & -3.27 & 0.07 & M20        \\
1RXS J175750.3+471634   & 5.45  & 0.05 & M20  & -0.705 & 0.040 & 29.09 & 0.35 & -3.36 & 0.42 & M20        \\
\textbf{LP 149-56}               & 6.17  & 0.06 & M20  & -0.661 & 0.041 & 28.84 & 0.03 & -3.23 & 0.04 & M20        \\
NGC 6475                &       &      &      & -0.660 & 0.100 & 29.41 & 0.33 & -2.89 & 0.36 & N16        \\
TYC 3059-299-1          & 6.46  & 0.06 & M20  & -0.643 & 0.041 & 29.30 & 0.02 & -3.16 & 0.02 & M20        \\
CD-72 1700A (GJ 1264)   & 6.67  & 0.07 & M20  & -0.629 & 0.041 & 29.31 & 0.17 & -2.92 & 0.20 & M22        \\
UCAC4 678-130516        & 6.85  & 0.07 & M20  & -0.618 & 0.042 & 29.04 & 0.11 & -3.50 & 0.13 & M20        \\
TYC 2703-706-1          & 7.80  & 0.21 & M20  & -0.559 & 0.044 & 29.14 & 0.06 & -3.34 & 0.07 & R, S       \\
\textbf{UCAC4 666-047939}        & 7.85  & 0.08 & M20  & -0.556 & 0.043 & 28.88 & 0.11 & -3.49 & 0.13 & M20        \\
EPIC 202059204          & 7.89  & 0.08 & M20  & -0.554 & 0.043 & 28.62 & 0.04 & -3.61 & 0.05 & M20        \\
1RXS J051125.0+265849   & 8.09  & 0.08 & M20  & -0.541 & 0.043 & 29.00 & 0.08 & -3.08 & 0.10 & M20        \\
UCAC4 729-006249        & 8.35  & 0.08 & M20  & -0.525 & 0.043 & 29.09 & 0.03 & -3.33 & 0.04 & M20        \\
TYC 3425-1133-1         & 8.35  & 0.08 & M20  & -0.525 & 0.043 & 28.92 & 0.02 & -3.33 & 0.02 & M20        \\
\textbf{2MASS J21154192+1746242} & 9.02  & 0.09 & M20  & -0.484 & 0.044 & 29.06 & 0.06 & -3.31 & 0.07 & M20        \\
UCAC4 643-115616        & 9.65  & 0.10 & M20  & -0.444 & 0.045 & 29.25 & 0.07 & -3.34 & 0.08 & M20        \\
CK Ari                  & 9.70  & 0.10 & M20  & -0.441 & 0.045 & 29.06 & 0.02 & -3.26 & 0.02 & M20        \\
UCAC4 654-053944        & 10.08 & 0.28 & M20  & -0.418 & 0.048 & 28.83 & 0.32 & -3.66 & 0.38 & C, R, S, X \\
UCAC4 669-046011        & 10.35 & 0.10 & M20  & -0.401 & 0.046 & 28.70 & 0.14 & -3.60 & 0.17 & M20        \\
TYC 3483-856-1          & 10.54 & 0.29 & M20  & -0.389 & 0.049 & 28.80 & 0.02 & -3.47 & 0.02 & R, S       \\
\textbf{EPIC 210741091}          & 10.95 & 0.11 & M20  & -0.364 & 0.047 & 28.88 & 0.02 & -3.54 & 0.02 & M20        \\
EPIC 210707811          & 11.77 & 0.12 & M20  & -0.313 & 0.048 & 28.79 & 0.04 & -3.34 & 0.05 & R, X       \\
GJ 3367                 & 12.05 & 0.12 & M20  & -0.295 & 0.048 & 28.79 & 0.57 & -3.65 & 0.68 & M22        \\
V2689 Ori               & 12.29 & 0.12 & M20  & -0.280 & 0.049 & 28.45 & 0.12 & -3.97 & 0.15 & M22, R     \\
M37                     &       &      &      & -0.280 & 0.050 & 29.11 & 0.22 & -3.13 & 0.24 & N16        \\
BD+16 2708 (GJ 569 A)   & 13.68 & 0.14 & M20  & -0.194 & 0.051 & 28.33 & 0.43 & -3.79 & 0.52 & C, R, X    \\
StKM 1-1215             & 13.91 & 0.38 & M20  & -0.180 & 0.056 & 28.48 & 0.15 & -4.03 & 0.18 & R, X       \\
DS Leo (GJ 410)         & 14.52 & 0.02 & 1A   & -0.142 & 0.052 & 28.35 & 0.14 & -3.99 & 0.17 & M22        \\
Hyaeds/Praesepe         &       &      &      & -0.137 & 0.070 & 28.76 & 0.47 & -3.58 & 0.49 & N16        \\
1RXS J133332.0+364153   & 15.36 & 0.15 & M20  & -0.090 & 0.054 & 28.82 & 0.22 & -3.47 & 0.26 & M20        \\
TYC 3142-1028-1         & 15.65 & 0.43 & M20  & -0.072 & 0.060 & 28.67 & 0.28 & -3.35 & 0.34 & R, X       \\
HD 11507                & 15.80 & 0.16 & M20  & -0.063 & 0.055 & 27.76 & 0.25 & -4.69 & 0.30 & M22        \\
UCAC4 650-018148        & 16.14 & 0.16 & M20  & -0.041 & 0.055 & 28.50 & 0.18 & -4.01 & 0.22 & M20        \\
StKM 1-1407             & 16.15 & 0.16 & M20  & -0.041 & 0.055 & 28.70 & 0.02 & -3.76 & 0.02 & M20        \\
GJ 3942                 & 16.30 & 0.45 & GA19 & -0.031 & 0.061 & 27.81 & 0.05 & -4.53 & 0.06 & F, G       \\
GJ 685                  & 16.30 & 0.16 & M20  & -0.031 & 0.055 & 27.82 & 0.30 & -4.52 & 0.36 & M20        \\
GJ 338 A                & 16.30 & 2.40 & GA19 & -0.031 & 0.159 & 27.99 & 0.40 & -4.53 & 0.48 & C, R       \\
G 115-72                & 16.61 & 0.17 & M20  & -0.012 & 0.056 & 28.49 & 0.08 & -3.97 & 0.10 & M20        \\
GJ 338 B                & 16.61 & 0.04 & GA19 & -0.012 & 0.055 & 27.96 & 0.34 & -4.47 & 0.41 & C, W       \\
GJ 694.2                & 17.30 & 0.17 & M20  & 0.031  & 0.057 & 27.98 & 0.15 & -4.36 & 0.18 & M20        \\
\textbf{EPIC 202059231}          & 17.43 & 0.17 & M20  & 0.039  & 0.057 & 28.46 & 0.04 & -3.66 & 0.05 & M20        \\
GJ 3822                 & 18.30 & 0.18 & M20  & 0.093  & 0.059 & 27.95 & 0.17 & -4.44 & 0.20 & M20        \\
GJ 49                   & 19.45 & 0.06 & 1R   & 0.164  & 0.060 & 27.57 & 0.04 & -4.70 & 0.07 & X          \\
GJ 606                  & 20.00 & 0.20 & M20  & 0.198  & 0.062 & 27.86 & 0.13 & -4.29 & 0.16 & M20        \\
1RXS J165257.4+255744   & 20.22 & 0.20 & M20  & 0.212  & 0.063 & 28.47 & 0.46 & -3.85 & 0.55 & R, X       \\
GJ 2                    & 21.20 & 0.21 & M20  & 0.273  & 0.065 & 27.60 & 0.13 & -4.62 & 0.16 & M20        \\
GJ 720A                 & 34.50 & 0.35 & M20  & 0.517  & 0.074 & 27.71 & 0.74 & -4.71 & 0.74 & M22        \\
GJ 47                   & 34.70 & 0.35 & M20  & 0.518  & 0.074 & 27.42 & 0.14 & -4.45 & 0.17 & M20        \\
HD 38529c               & 35.90 & 0.36 & M20  & 0.530  & 0.076 & 27.05 & 0.15 & -4.97 & 0.18 & M20        \\
GJ 740                  & 36.40 & 0.36 & M20  & 0.534  & 0.077 & 27.45 & 0.14 & -4.95 & 0.17 & M20        \\
\textbf{GJ 832}                  & 38.10 & 0.10 & 1S   & 0.550  & 0.080 & 26.59 & 0.28 & -5.38 & 0.34 & R, S, X    \\
EPIC 201675315          & 38.13 & 0.38 & M20  & 0.550  & 0.081 & 26.97 & 0.03 & -5.38 & 0.04 & M20        \\
HD 285968 (GJ 176)      & 40.13 & 1.13 & 1R   & 0.569  & 0.085 & 27.25 & 0.33 & -4.88 & 0.40 & C, R       \\
\textbf{BD-18 359}               & 44.51 & 0.45 & M20  & 0.610  & 0.093 & 27.45 & 0.67 & -4.65 & 0.67 & R, M22     \\
LHS 373 (GJ 552)        & 44.90 & 1.20 & 1R   & 0.614  & 0.095 & 27.50 & 0.19 & -4.68 & 0.23 & M20        \\
GJ 15 A                 & 44.95 & 0.55 & 1R   & 0.614  & 0.095 & 26.41 & 0.33 & -5.52 & 0.40 & M20        \\
GJ 752 A                & 46.60 & 0.20 & 1H   & 0.630  & 0.098 & 26.91 & 0.17 & -5.12 & 0.21 & R          \\
GJ 625                  & 51.22 & 0.21 & 1R   & 0.673  & 0.108 & 26.82 & 0.17 & -4.82 & 0.22 & R          \\
BR Psc (GJ 908)         & 53.70 & 0.80 & 1A   & 0.696  & 0.114 & 26.47 & 0.11 & -5.48 & 0.13 & M20        \\
GJ 476                  & 55.00 & 5.50 & SM18 & 0.708  & 0.127 & 27.05 & 0.15 & -4.82 & 0.22 & GA19       \\
HD 95735b (GJ 411)      & 56.15 & 0.27 & DA19 & 0.719  & 0.119 & 27.07 & 0.19 & -5.05 & 0.23 & M20, X     \\
GJ 1                    & 56.80 & 5.60 & SM17 & 0.725  & 0.131 & 26.71 & 0.28 & -5.27 & 0.34 & R          \\
GJ 625                  & 60.70 & 0.30 & 1R   & 0.761  & 0.129 & 26.90 & 0.14 & -4.82 & 0.17 & M20        \\
\textbf{GJ 433}                  & 72.60 & 0.50 & 1F   & 0.873  & 0.156 & 26.51 & 0.31 & -5.58 & 0.37 & S, M22     \\
\enddata
\tablecomments{\footnotesize{$^1$ Measured in this study using data from: A = \textit{APT}, F = \citet{2020ApJS..246...11F}, M = \textit{MEarth}, R = \textit{RCT}, S = \textit{Skynet}, Z = \textit{ZTF})  -- H11: \citet{2011AJ....141..166H}  -- SM15: \citet{2015MNRAS.452.2745S}  --  N16: \citet{2016ApJ...830...44N}  -- D19: \citet{2019AA...625A..17D}  -- GA19: \citet{2019AA...624A..27G}  -- M20: \citet{2020AA...638A..20M}  -- M22: \citet{2022AN....34320049M}  
\newline For X-ray sources: -- C = \textit{Chandra}  -- R = \textit{ROSAT}  -- S = \textit{Swift}  -- X = \textit{XMM}
}}
\end{deluxetable*}
\end{longrotatetable}

\begin{longrotatetable}
\begin{deluxetable*}{lcccccccccc}
\tablecaption{Mid-Late M Dwarf X-ray Data\label{table:midlatexraydata}}
\tablehead{\colhead{Star Name}  & \colhead{$P_{\rm{rot}}$}  & \colhead{err}
           & \colhead{$P_{\rm{rot}}$ src}  & \colhead{$\log$ Age}  & \colhead{err}  
           & \colhead{$\log(\langle L_{\rm X}\rangle$)}  
           & \colhead{err}  
           & \colhead{$\log(\langle L_{\rm X}\rangle$ / $L_{\rm bol})$}
           & \colhead{err}  & \colhead{X-ray src}}
\startdata
Pleiades/NGC   2516      &        &      &         & -0.870 & 0.120 & 28.98 & 0.31 & -2.92 & 0.35 & N16        \\
LP 263-64                & 2.06   & 0.02 & M20     & -0.774 & 0.019 & 28.70 & 0.10 & -2.95 & 0.12 & M22        \\
1RXS J175537.7+485740    & 2.24   & 0.02 & M20     & -0.764 & 0.019 & 28.57 & 0.07 & -3.08 & 0.08 & M20        \\
BD+20 2465 (AD Leo)      & 2.24   & 0.02 & 1R, M20 & -0.764 & 0.019 & 28.81 & 0.05 & -3.14 & 0.06 & M22        \\
2MASS J16243140+4549570  & 2.36   & 0.02 & M20     & -0.758 & 0.019 & 28.60 & 0.55 & -2.86 & 0.66 & M20        \\
LP 302-37                & 2.60   & 0.03 & M20     & -0.744 & 0.019 & 29.26 & 0.08 & -2.98 & 0.10 & M20        \\
\textbf{1RXS J031028.8+285952}    & 2.87   & 0.03 & M20     & -0.729 & 0.019 & 28.40 & 0.13 & -3.38 & 0.16 & M20        \\
Ross 154 (GJ 729)        & 2.87   & 0.03 & M20     & -0.729 & 0.019 & 27.91 & 0.20 & -3.28 & 0.24 & C, R, X    \\
\textbf{LP 195-34}                & 3.39   & 0.03 & M20     & -0.700 & 0.019 & 28.99 & 0.70 & -2.70 & 0.84 & M20, S     \\
1RXS J222653.9+354610    & 3.55   & 0.04 & M20     & -0.691 & 0.019 & 28.70 & 0.10 & -2.98 & 0.12 & M20        \\
2MASS J16182491+4437110  & 4.07   & 0.04 & M20     & -0.662 & 0.019 & 28.44 & 0.08 & -2.99 & 0.10 & M20        \\
NGC 6475                 &        &      &         & -0.660 & 0.100 & 28.91 & 0.15 & -3.10 & 0.17 & N16        \\
1RXS J174409.9+473444    & 4.45   & 0.04 & M20     & -0.640 & 0.019 & 28.46 & 0.08 & -3.26 & 0.10 & M20, S     \\
UCAC4 610-011032         & 5.85   & 0.06 & M20     & -0.562 & 0.020 & 29.39 & 0.02 & -2.79 & 0.02 & M20        \\
HAT 214-03527            & 6.55   & 0.07 & M20     & -0.523 & 0.020 & 28.50 & 0.13 & -3.34 & 0.16 & M20        \\
2MASS J22170696+3627164  & 7.04   & 0.07 & M20     & -0.495 & 0.021 & 28.62 & 0.08 & -3.11 & 0.10 & M20        \\
CW UMa                   & 7.78   & 0.08 & M20     & -0.454 & 0.021 & 28.61 & 0.18 & -3.04 & 0.22 & M22        \\
BL Lyn                   & 7.85   & 0.08 & 1T, M22 & -0.450 & 0.021 & 28.88 & 0.25 & -3.17 & 0.30 & M22        \\
2MASS J08445708+4640261  & 8.81   & 0.09 & M20     & -0.396 & 0.022 & 29.16 & 0.03 & -2.63 & 0.04 & M20        \\
LP 197-37                & 9.36   & 0.09 & M20     & -0.365 & 0.022 & 28.35 & 0.07 & -3.50 & 0.08 & M20        \\
\textbf{G 176-13}                 & 10.75  & 0.11 & M20     & -0.287 & 0.023 & 29.17 & 0.07 & -2.76 & 0.08 & M20        \\
M37                      &        &      &         & -0.284 & 0.050 & 29.11 & 0.47 & -3.14 & 0.40 & N16        \\
\textbf{HAT 170-05945}            & 10.86  & 0.11 & M20     & -0.281 & 0.023 & 28.83 & 0.13 & -3.28 & 0.16 & M20        \\
HAT 136-02868            & 11.99  & 0.12 & M20     & -0.217 & 0.024 & 28.91 & 0.08 & -3.02 & 0.10 & C, R       \\
\textbf{HAT 216-04245}            & 12.49  & 0.12 & M20     & -0.189 & 0.025 & 28.70 & 0.14 & -3.14 & 0.17 & R, X       \\
1RXS J040731.6+340525    & 12.54  & 0.13 & M20     & -0.187 & 0.025 & 28.90 & 0.10 & -3.10 & 0.12 & M20        \\
Hyades/Praesepe          &        &      &         & -0.155 & 0.069 & 28.41 & 0.49 & -3.37 & 0.53 & N16        \\
PM J04310+3647           & 13.26  & 0.13 & M20     & -0.146 & 0.025 & 28.37 & 0.08 & -3.45 & 0.10 & M20        \\
G 123-35                 & 13.59  & 0.14 & M20     & -0.128 & 0.026 & 28.65 & 0.06 & -3.02 & 0.07 & M20        \\
EPIC 210434976           & 2.55   & 0.03 & M20     & -0.097 & 0.031 & 28.81 & 0.46 & -2.58 & 0.55 & F, S       \\
YZ CMi (GJ 285)          & 2.78   & 0.03 & M20     & -0.092 & 0.031 & 28.50 & 0.01 & -3.15 & 0.01 & M20        \\
EPIC 204957517           & 2.82   & 0.03 & M20     & -0.091 & 0.031 & 28.61 & 0.10 & -2.84 & 0.12 & M22        \\
\textbf{LP 331-57 A}              & 14.54  & 0.15 & M20     & -0.074 & 0.027 & 28.53 & 0.05 & -3.21 & 0.06 & M20        \\
\textbf{UCAC4 617-130729}         & 14.75  & 0.15 & M20     & -0.063 & 0.027 & 29.11 & 0.51 & -3.40 & 0.61 & M20        \\
HAT 122-01032            & 4.38   & 0.04 & M20     & -0.052 & 0.031 & 28.91 & 0.01 & -2.94 & 0.01 & M20        \\
\textbf{UCAC3 242-88245}          & 15.10  & 0.15 & M20     & -0.043 & 0.027 & 28.61 & 0.09 & -3.22 & 0.11 & M22        \\
EPIC 206262336           & 9.71   & 0.10 & M20     & 0.082  & 0.035 & 28.26 & 0.03 & -2.95 & 0.04 & M20        \\
HG 8-1                   & 17.57  & 0.18 & M20     & 0.096  & 0.030 & 28.20 & 0.21 & -3.61 & 0.25 & M20        \\
EPIC 201482319           & 11.68  & 0.12 & M20     & 0.132  & 0.037 & 28.50 & 0.04 & -3.19 & 0.05 & M20        \\
G 74-25                  & 15.29  & 0.15 & M20     & 0.222  & 0.041 & 28.47 & 0.05 & -3.30 & 0.06 & R, X       \\
UCAC4 619-055248         & 19.84  & 0.20 & M20     & 0.223  & 0.032 & 28.41 & 0.20 & -3.39 & 0.24 & F          \\
Ross 868 (GJ 669 A)      & 20.95  & 0.03 & 1R      & 0.285  & 0.031 & 28.51 & 0.29 & -3.33 & 0.35 & C, R       \\
\textbf{UCAC4 692-060564}         & 21.52  & 0.22 & M20     & 0.317  & 0.034 & 28.63 & 0.08 & -3.38 & 0.10 & M20        \\
GJ 4334                  & 23.54  & 0.24 & N16     & 0.449  & 0.027 & 28.46 & 0.24 & -3.01 & 0.29 & R, X       \\
\textbf{CD-40 5404B (GJ 358)}     & 25.26  & 0.70 & 1S      & 0.472  & 0.031 & 28.05 & 0.15 & -3.90 & 0.18 & M20        \\
LP 167-71                & 27.82  & 0.28 & M20     & 0.483  & 0.032 & 28.88 & 0.03 & -3.07 & 0.04 & M20        \\
2MASS J00240376+2626299  & 29.84  & 0.30 & M20     & 0.491  & 0.034 & 27.87 & 0.13 & -3.53 & 0.16 & M20        \\
1RXS J075235.5+410055    & 30.33  & 0.30 & M20     & 0.493  & 0.034 & 28.33 & 0.13 & -3.40 & 0.16 & M20        \\
1RXS J174158.2+475403    & 32.23  & 0.32 & M20     & 0.500  & 0.036 & 28.35 & 0.12 & -3.31 & 0.14 & M20        \\
GJ 674                   & 33.29  & 0.33 & AD17    & 0.505  & 0.037 & 27.52 & 0.17 & -4.26 & 0.20 & R, S, X    \\
2MASS J06170531+8353354  & 40.75  & 0.10 & 1M      & 0.535  & 0.043 & 27.38 & 0.27 & -4.29 & 0.32 & W, R       \\
2MASS J20125995+0112584  & 41.15  & 1.23 & 13      & 0.537  & 0.044 & 27.95 & 0.49 & -3.29 & 0.59 & C, R       \\
GJ 752 A                 & 46.60  & 0.20 & 1H      & 0.559  & 0.049 & 26.91 & 0.17 & -5.12 & 0.21 & R          \\
GJ 588                   & 51.70  & 0.10 & 1H      & 0.579  & 0.055 & 26.92 & 0.19 & -4.67 & 0.24 & R          \\
2MASS J06575703+6219197  & 54.50  & 0.55 & M20     & 0.591  & 0.057 & 27.83 & 0.07 & -3.56 & 0.08 & M20        \\
2MASS J16492028+5101177  & 54.71  & 0.55 & M20     & 0.591  & 0.058 & 26.89 & 0.43 & -4.77 & 0.52 & M20        \\
GJ 163                   & 56.04  & 2.47 & 1M      & 0.597  & 0.060 & 26.87 & 0.15 & -5.09 & 0.18 & Fo22       \\
2MASS J05011802+2237015  & 70.67  & 0.71 & M20     & 0.656  & 0.076 & 27.40 & 0.12 & -3.59 & 0.14 & M20        \\
2MASS J19310458-0306186  & 70.92  & 0.71 & M20     & 0.657  & 0.076 & 26.82 & 0.11 & -3.94 & 0.13 & M20        \\
\textbf{GJ 436}                   & 71.40  & 0.40 & 1R      & 0.659  & 0.076 & 26.85 & 0.46 & -5.11 & 0.55 & C, R, X    \\
2MASS J13505181+3644168  & 72.18  & 0.72 & M20     & 0.662  & 0.077 & 26.84 & 0.30 & -4.62 & 0.36 & M20        \\
2MASS J17311725+8205198  & 79.22  & 0.79 & M20     & 0.691  & 0.085 & 27.40 & 0.09 & -3.70 & 0.11 & M20        \\
GJ 3253                  & 81.55  & 0.82 & M20     & 0.700  & 0.088 & 26.94 & 0.04 & -4.16 & 0.05 & M20        \\
2MASS J03103891+2540535  & 82.83  & 0.83 & M20     & 0.706  & 0.089 & 26.95 & 0.30 & -4.45 & 0.36 & M20        \\
GJ 357                   & 86.10  & 0.90 & 1H      & 0.719  & 0.093 & 26.94 & 0.25 & -4.70 & 0.30 & X          \\
G 184-31                 & 86.25  & 0.86 & M20     & 0.719  & 0.093 & 26.72 & 0.09 & -4.41 & 0.11 & M20        \\
EPIC 206019392 (G 876)   & 87.95  & 0.61 & 1A      & 0.726  & 0.095 & 26.34 & 0.10 & -5.38 & 0.12 & F, X       \\
GJ 551                   & 88.98  & 0.01 & 1S      & 0.730  & 0.096 & 26.52 & 0.39 & -4.27 & 0.47 & W          \\
2MASS J18551471-7115026  & 90.69  & 0.91 & M20     & 0.737  & 0.098 & 26.63 & 0.30 & -4.95 & 0.36 & M20        \\
2MASS J14211512-0107199  & 91.41  & 0.91 & M20     & 0.740  & 0.099 & 26.27 & 0.30 & -5.13 & 0.36 & M20        \\
2MASS J05015746-0656459  & 92.15  & 2.76 & M20     & 0.743  & 0.101 & 27.32 & 0.08 & -3.82 & 0.10 & C, R       \\
2MASS J19194119-5955194  & 92.69  & 0.93 & M20     & 0.745  & 0.100 & 26.36 & 0.30 & -5.05 & 0.36 & M20        \\
2MASS J23351050-0223214  & 100.40 & 1.00 & M20     & 0.777  & 0.110 & 26.05 & 0.43 & -4.71 & 0.52 & M20        \\
\textbf{GJ 667 C}                 & 103.10 & 0.60 & 1H      & 0.788  & 0.112 & 26.69 & 0.28 & -5.06 & 0.34 & C, R       \\
2MASS J20403364+1529572  & 104.60 & 1.05 & M20     & 0.794  & 0.114 & 26.60 & 0.13 & -4.62 & 0.16 & M20        \\
GJ 628                   & 108.70 & 1.50 & 1R      & 0.810  & 0.119 & 26.40 & 0.55 & -5.21 & 0.66 & C, R       \\
EPIC 201518346           & 112.83 & 1.13 & M20     & 0.827  & 0.124 & 26.74 & 0.03 & -4.42 & 0.04 & M20        \\
EPIC 205913009 (GJ 1265) & 116.40 & 1.16 & M20     & 0.841  & 0.129 & 26.03 & 0.03 & -5.11 & 0.04 & M20        \\
2MASS J11505787+4822395  & 117.50 & 1.18 & M20     & 0.846  & 0.129 & 26.20 & 0.30 & -4.92 & 0.36 & M20        \\
GJ 447                   & 124.95 & 1.90 & 1R      & 0.876  & 0.139 & 26.71 & 0.36 & -4.88 & 0.43 & M22        \\
2MASS J19204795-4533283  & 133.90 & 1.34 & M20     & 0.912  & 0.149 & 26.51 & 0.30 & -4.66 & 0.36 & M20        \\
GJ 699                   & 149.50 & 0.60 & 1AR,3   & 0.968  & 0.069 & 25.53 & 0.54 & -5.57 & 0.65 & F20, R     \\
GJ 273                   & 160.83 & 2.48 & 1R      & 0.968  & 0.069 & 26.36 & 0.32 & -5.23 & 0.38 & R          \\
GJ 581                   & 147.80 & 0.60 & 1AR     & 0.968  & 0.069 & 25.73 & 0.55 & -5.94 & 0.66 & C, S       \\
GJ 191                   & 153.20 & 3.70 & 1S      & 1.021  & 0.046 & 26.46 & 0.50 & -5.23 & 0.60 & C, R, S, X \\
\enddata
\tablecomments{\footnotesize{$^1$ Measured in this study using data from: A = \textit{APT}, F = \citet{2020ApJS..246...11F}, M = \textit{MEarth}, R = \textit{RCT}, S = \textit{Skynet}, Z = \textit{ZTF})  --  $^*$ Age calculated using the M4+ young track relationship -- H11: \citet{2011AJ....141..166H}  -- N16: \citet{2016ApJ...830...44N}  -- PL16: \citet{2016ApJ...824..102L}  --  AD17: \citet{2017AA...600A..13A}  -- W18: \citet{2018MNRAS.479.2351W}  -- N18:  \citet{2018AJ....156..217N}  -- D19: \citet{2019AA...625A..17D}  -- T19: \citet{2019MNRAS.488.5145T}  -- M20: \citet{2020AA...638A..20M}  -- M22: \citet{2022AN....34320049M}  --  F20: \citet{2020AJ....160..237F}  --  F22:  \citet{2022AA...664A.105F}  --  Fo22: \citet{2022AA...661A..23F}  -- C = \textit{Chandra}  -- R = \textit{ROSAT}  -- S = \textit{Swift}  -- X = \textit{XMM}
}}
\end{deluxetable*}
\end{longrotatetable}

\begin{longrotatetable}
\begin{deluxetable*}{lccccccc}
\tablecaption{Early M Dwarf Ca {\textsc{ii}} Data\label{table:earlycaiidata}}
\tablehead{\colhead{Star Name}  & \colhead{$P_{\rm{rot}}$}  & \colhead{err}
           & \colhead{$P_{\rm{rot}}$ src}  & \colhead{$\log$ Age}  & \colhead{err}  
           & \colhead{$\log~\langle R_{\rm HK}\rangle$}  
           & \colhead{err}}
\startdata
GJ103                  & 1.56           & 0.02 & AD17 & -0.947 & 0.038 & -3.877 & 0.100 \\
Pleiades               &      &      & -0.870 & 0.120 & -4.170 & 0.290 \\
GJ 494                  & 2.89           & 0.03 & AD17 & -0.864 & 0.039 & -3.930 & 0.130 \\
GJ 9520                 & 3.37           & 0.03 & AD17 & -0.834 & 0.039 & -3.978 & 0.750 \\
GJ 867A                 & 4.23           & 0.04 & AD17 & -0.781 & 0.039 & -4.149 & 0.120 \\
GJ 182                  & 4.41           & 0.04 & AD17 & -0.770 & 0.040 & -3.865 & 0.450 \\
GJ 803                  & 4.85           & 0.05 & AD17 & -0.743 & 0.040 & -4.022 & 0.210 \\
GJ 1264                 & 6.67           & 0.07 & AD17 & -0.629 & 0.041 & -3.994 & 0.190 \\
GJ 208                  & 11.49          & 0.36 & M22  & -0.330 & 0.052 & -4.140 & 0.625 \\
\textbf{GJ 3685}                 & 11.60          & 0.12 & P23  & -0.323 & 0.048 & -4.229 & 0.402 \\
GJ 3367                 & 12.05          & 0.12 & AD17 & -0.295 & 0.048 & -4.088 & 1.157 \\
DS Leo(GJ 410)           & 14.52          & 0.02 & 1A   & -0.142 & 0.052 & -4.123 & 0.352 \\
Hyades                 &                &      &      & -0.137 & 0.070 & -4.510 & 0.360 \\
GJ 3942                 & 16.3           & 0.1  & SM18 & -0.031 & 0.055 & -4.550 & 0.100 \\
GJ 338A                 & 16.3           & 2.4  & SM18 & -0.031 & 0.159 & -4.455 & 0.100 \\
GJ 338B                 & 16.61          & 0.04 & R23  & -0.012 & 0.055 & -4.470 & 0.100 \\
GJ 373                  & 17.18          & 1    & 1T   & 0.023  & 0.084 & -4.423 & 0.057 \\
GJ 694.2                & 17.3           & 0.1  & SM18 & 0.031  & 0.057 & -4.460 & 0.150 \\
GJ 21                   & 17.4           & 1.1  & SM18 & 0.037  & 0.089 & -4.720 & 0.030 \\
GJ 3822                 & 18.3           & 0.1  & SM18 & 0.093  & 0.058 & -4.700 & 0.040 \\
GJ 49                   & 19.45          & 0.06 & 1R   & 0.164  & 0.060 & -4.401 & 0.588 \\
GJ 606                  & 20             & 2    & SM18 & 0.198  & 0.138 & -4.890 & 0.060 \\
GJ 3218                 & 21.10          & 0.21 & R23  & 0.267  & 0.065 & -4.517 & 0.276 \\
GJ 2                    & 21.2           & 0.5  & GA19 & 0.273  & 0.071 & -4.622 & 0.540 \\
GJ 382                  & 21.20          & 0.10 & SM16 & 0.273  & 0.064 & -4.655 & 0.247 \\
GJ 649                  & 21.34          & 0.29 & 1R   & 0.282  & 0.066 & -4.621 & 0.695 \\
GJ 3470                 & 21.54          & 0.49 & K19  & 0.294  & 0.071 & -4.883 & 0.210 \\
GJ 685                  & 21.83          & 3.76 & 1A   & 0.312  & 0.242 & -4.596 & 0.036 \\
GJ 450                  & 22.8           & 1    & DA19 & 0.372  & 0.091 & -4.622 & 0.392 \\
GJ 9404                 & 23.2           & 0.1  & SM18 & 0.403  & 0.049 & -4.710 & 0.040 \\
GJ 846                  & 26.30          & 5.60 & SM15 & 0.440  & 0.080 & -4.820 & 0.080 \\
GJ 4057                 & 26.7           & 0.1  & SM18 & 0.444  & 0.061 & -5.030 & 0.040 \\
GJ 4306                 & 27             & 2.5  & SM18 & 0.447  & 0.066 & -4.840 & 0.040 \\
GJ 229                  & 27.3           & 0.1  & SM16 & 0.450  & 0.062 & -4.773 & 0.275 \\
GJ 740                  & 27.51          & 0.07 & 1AS  & 0.451  & 0.062 & -4.578 & 0.367 \\
GJ 514                  & 30             & 0.9  & SM17 & 0.475  & 0.067 & -4.879 & 0.127 \\
GJ 1030                 & 32             & 3    & SM18 & 0.493  & 0.075 & -4.840 & 0.060 \\
GJ 162                  & 32.4           & 1.6  & SM18 & 0.497  & 0.071 & -4.950 & 0.040 \\
GJ 3998                 & 33.6           & 3.6  & SM18 & 0.508  & 0.079 & -5.010 & 0.050 \\
GJ 205                  & 33.72          & 0.13 & 1R   & 0.509  & 0.072 & -4.743 & 0.393 \\
GJ 393                  & 34.15          & 0.34 & AD17 & 0.513  & 0.074 & -5.031 & 0.106 \\
GJ 720A                 & 34.5           & 4.7  & SM18 & 0.517  & 0.086 & -4.859 & 0.170 \\
GJ 47                   & 34.7           & 0.1  & SM18 & 0.518  & 0.074 & -4.910 & 0.060 \\
GJ 9689                 & 35.7           & 0.2  & SM18 & 0.528  & 0.076 & -4.820 & 0.050 \\
GJ 548A                 & 36.6           & 0.1  & SM18 & 0.536  & 0.078 & -4.780 & 0.030 \\
GJ 3997                 & 37             & 13   & SM18 & 0.540  & 0.144 & -4.780 & 0.030 \\
GJ 880                  & 37.5           & 0.38 & AD17 & 0.544  & 0.079 & -4.741 & 0.585 \\
GJ 832                 & 38.1           & 0.11 & 1S   & 0.550  & 0.079 & -5.148 & 0.172 \\
GJ 2066                 & 40.7           & 0.41 & AD17 & 0.574  & 0.086 & -5.220 & 0.077 \\
GJ 676A                 & 41.2           & 3.8  & SM15 & 0.579  & 0.094 & -5.033 & 0.110 \\
GJ 156.1A               & 41.2           & 6.3  & SM18 & 0.579  & 0.105 & -5.040 & 0.040 \\
GJ 3138                 & 42             & 0.42 & AD17 & 0.587  & 0.088 & -4.855 & 0.126 \\
GJ 552                  & 43.5           & 1.5  & SM18 & 0.601  & 0.092 & -5.070 & 0.040 \\
GJ 15A                  & 44.99          & 3.4  & 1R   & 0.615  & 0.099 & -5.270 & 0.100 \\
GJ 184                  & 45             & 0.1  & SM18 & 0.615  & 0.095 & -5.090 & 0.040 \\
GJ 412A                 & 46.4           & 0.2  & SM18 & 0.628  & 0.097 & -5.460 & 0.150 \\
GJ 9440                 & 48             & 4.8  & SM18 & 0.643  & 0.111 & -5.250 & 0.040 \\
GJ 654                  & 49             & 0.49 & AD17 & 0.652  & 0.103 & -5.266 & 0.093 \\
GJ 521A                 & 49.5           & 3.5  & SM18 & 0.657  & 0.109 & -5.360 & 0.050 \\
GJ 536                  & 49.8           & 0.1  & 1H   & 0.660  & 0.104 & -5.059 & 0.130 \\
V*BR Psc (GJ 908)      & 49.9           & 3.5  & SM18 & 0.661  & 0.110 & -5.380 & 0.060 \\
GJ 119A                 & 51.2           & 4.4  & SM18 & 0.673  & 0.115 & -4.970 & 0.040 \\
GJ 625                  & 51.22          & 0.21 & 1R   & 0.673  & 0.108 & -5.333 & 1.273 \\
GJ 526                  & 52.3           & 0.4  & 1H   & 0.683  & 0.110 & -5.113 & 0.089 \\
\textbf{GJ 476}                  & 55             & 5.5  & SM18 & 0.708  & 0.127 & -5.210 & 0.050 \\
GJ 411                  & 56.15          & 0.27 & DA19 & 0.719  & 0.119 & -5.371 & 0.100 \\
GJ 1                    & 56.8           & 5.6  & SM17 & 0.725  & 0.131 & -5.431 & 0.121 \\
\textbf{GJ 433}                  & 72.6           & 0.5  & 1F   & 0.873  & 0.156 & -5.289 & 0.384 \\
\enddata
\tablecomments{\footnotesize{$^1$ Measured in this study using data from: A = \textit{APT}, F = \citet{2020ApJS..246...11F}, H = \citet{2020AA...636A..74T} ,M = \textit{MEarth}, R = \textit{RCT}, S = \textit{Skynet}, Z = \textit{ZTF})  --  KS13: \citet{2013AcA....63...53K}  -- SM15: \citet{2015MNRAS.452.2745S}  -- SM16: \citet{2016AA...595A..12S}   -- N16: \citet{2016ApJ...821...93N}  -- SM17:  \citet{2017MNRAS.468.4772S}  --  AD17: \citet{2017AA...600A..13A}  -- N18:  \citet{2018AJ....156..217N}  -- SM18:  \citet{2018AA...612A..89S}  -- DA19: \citet{2019AA...621A.126D}  --  B22: \citet{2022ApJ...929...80B}
\newline ** Each star's $\log~\langle R_{\rm HK}\rangle$  value was determined using the combined data (depending on availability) from \citet{2013AA...551L...8P,2015MNRAS.452.2745S,2016AA...595A..12S,2017AA...600A..13A,2017MNRAS.468.4772S,2018AA...616A.108B,2018AA...612A..89S,2020AJ....160..269M,2021AA...652A.116P,2022ApJ...929...80B,2023AA...672A..37I,2023AA...671A.162M}; the Pleiades and Hyades values were obtained from \citet{2018MNRAS.476..908F}.
}}
\end{deluxetable*}
\end{longrotatetable}

\begin{longrotatetable}
\begin{deluxetable*}{lccccccc}
\tablecaption{Mid-Late M Dwarf Ca {\textsc{ii}} Data\label{table:midlatecaiidata}}
\tablehead{\colhead{Star Name}  & \colhead{$P_{\rm{rot}}$}  & \colhead{err}
           & \colhead{$P_{\rm{rot}}$ src}  & \colhead{$\log$ Age (Gyr)}  & \colhead{err}  
           & \colhead{$\log~\langle R_{\rm HK}\rangle$}  
           & \colhead{err}}
\startdata
Pleiades         &        &      &      & -0.870 & 0.120 & -4.201 & 0.260 \\
LP 831-35        & 2.08   & 0.02 & N18  & -0.773 & 0.019 & -4.234 & 0.002 \\
AD Leo (GJ 388)  & 2.24   & 0.02 & 1R   & -0.764 & 0.019 & -4.191 & 0.070 \\
LP 805-1         & 2.81   & 0.03 & N18  & -0.732 & 0.019 & -4.390 & 0.004 \\
GJ 729           & 2.87   & 0.03 & AD17 & -0.729 & 0.019 & -4.405 & 0.510 \\
LP 709-18        & 3.15   & 0.03 & N18  & -0.713 & 0.019 & -4.540 & 0.002 \\
UCAC3 151-377752 & 6.72   & 0.07 & N18  & -0.513 & 0.021 & -4.491 & 0.002 \\
EPIC 201482319   & 11.68  & 0.12 & M20  & -0.235 & 0.024 & -4.228 & 0.402 \\
Hyades           &        &      &      & -0.137 & 0.070 & -4.330 & 0.480 \\
LP 735-29        & 2.10   & 0.02 & N18  & -0.109 & 0.031 & -4.405 & 0.003 \\
G 70-46          & 2.35   & 0.02 & N16  & -0.103 & 0.031 & -4.473 & 0.002 \\
GJ 285           & 2.78   & 0.03 & AD17 & -0.092 & 0.031 & -4.151 & 0.510 \\
LP 851-399       & 3.07   & 0.03 & N18  & -0.084 & 0.031 & -4.295 & 0.003 \\
GJ 3127          & 14.79  & 0.15 & N18  & -0.060 & 0.027 & -4.791 & 0.003 \\
G 152-1          & 4.38   & 0.04 & N18  & -0.052 & 0.031 & -4.621 & 0.002 \\
GJ 4274          & 4.41   & 0.30 & 1T   & -0.051 & 0.032 & -4.350 & 0.250 \\
LHS 1358         & 4.73   & 0.05 & N16  & -0.043 & 0.031 & -4.635 & 0.001 \\
SCR J1107-3420B  & 7.61   & 0.08 & N18  & 0.030  & 0.033 & -4.225 & 0.003 \\
BL Lyn           & 16.40  & 0.16 & R23  & 0.030  & 0.028 & -4.280 & 0.049 \\
GJ 4071          & 8.16   & 0.25 & 1T   & 0.043  & 0.034 & -4.292 & 0.221 \\
G 151-64         & 8.43   & 0.08 & N18  & 0.050  & 0.034 & -4.202 & 0.002 \\
GJ 669 A         & 20.95  & 0.03 & 1R   & 0.285  & 0.031 & -4.570 & 0.400 \\
GJ 479           & 23.75  & 0.24 & AD17 & 0.454  & 0.028 & -4.766 & 0.284 \\
GJ 362           & 24.87  & 0.04 & 1R   & 0.471  & 0.031 & -4.493 & 0.436 \\
LP 740-10        & 25.01  & 0.25 & N18  & 0.471  & 0.031 & -4.764 & 0.003 \\
GJ 358           & 25.26  & 0.70 & 1S   & 0.472  & 0.031 & -4.655 & 0.077 \\
LHS 5273         & 30.40  & 0.30 & N18  & 0.493  & 0.034 & -4.609 & 0.004 \\
GJ 674           & 33.29  & 0.33 & AD17 & 0.505  & 0.037 & -4.978 & 0.210 \\
\textbf{GJ 832}           & 34.78  & 2.82 & 1S   & 0.511  & 0.040 & -5.148 & 0.172 \\
GJ 176           & 40.14  & 0.57 & 1R   & 0.533  & 0.043 & -4.949 & 0.091 \\
GJ 3293          & 41.00  & 0.40 & AD17 & 0.536  & 0.044 & -5.114 & 0.070 \\
LP 734-34        & 42.99  & 0.43 & N18  & 0.544  & 0.045 & -4.689 & 0.003 \\
GJ 752A          & 46.60  & 0.20 & 1H   & 0.559  & 0.049 & -5.113 & 0.102 \\
GJ 889 B         & 51.20  & 0.51 & N18  & 0.577  & 0.054 & -5.278 & 0.008 \\
\textbf{GJ 588}           & 51.70  & 0.52 & 1H   & 0.579  & 0.055 & -5.143 & 0.403 \\
GJ 877           & 52.80  & 0.53 & SM16 & 0.584  & 0.056 & -5.180 & 0.293 \\
GJ 1088          & 53.74  & 0.54 & N18  & 0.587  & 0.057 & -5.251 & 0.014 \\
GJ 163           & 56.04  & 2.47 & 1M   & 0.597  & 0.060 & -5.422 & 0.130 \\
GJ 618A          & 56.52  & 0.60 & AD17 & 0.599  & 0.060 & -5.401 & 0.070 \\
WT 244           & 66.45  & 0.66 & N18  & 0.639  & 0.071 & -5.603 & 0.015 \\
LP 816-60        & 67.60  & 0.68 & SM16 & 0.644  & 0.072 & -5.116 & 0.488 \\
GJ 849           & 71.05  & 0.50 & AD17 & 0.657  & 0.076 & -5.054 & 0.540 \\
\textbf{GJ 436}           & 71.40  & 0.40 & 1R   & 0.659  & 0.076 & -5.269 & 0.599 \\
GJ 1148          & 71.50  & 5.10 & DA19 & 0.659  & 0.079 & -5.336 & 1.030 \\
L 154-205        & 73.14  & 0.73 & N18  & 0.666  & 0.079 & -4.874 & 0.005 \\
GJ 12            & 78.50  & 0.80 & AD17 & 0.688  & 0.084 & -5.368 & 0.070 \\
LHS 610          & 78.80  & 0.80 & AD17 & 0.689  & 0.085 & -5.375 & 0.070 \\
LP 780-32        & 79.15  & 0.79 & N18  & 0.691  & 0.085 & -5.252 & 0.007 \\
L 50-78          & 80.97  & 0.81 & N18  & 0.698  & 0.087 & -5.388 & 0.007 \\
GJ 4132          & 84.99  & 0.85 & N18  & 0.714  & 0.092 & -5.653 & 0.008 \\
\textbf{GJ 357}           & 86.10  & 0.90 & 1H   & 0.719  & 0.093 & -5.430 & 0.710 \\
GJ 876           & 87.95  & 0.61 & 1A   & 0.726  & 0.095 & -5.258 & 0.260 \\
GJ 3323          & 88.50  & 0.89 & KS12 & 0.728  & 0.095 & -4.747 & 0.436 \\
GJ 551           & 88.98  & 0.01 & 1S   & 0.730  & 0.096 & -5.134 & 0.243 \\
GJ 876           & 91.00  & 0.90 & AD17 & 0.738  & 0.099 & -5.496 & 0.070 \\
GJ 3843          & 91.43  & 0.91 & N18  & 0.740  & 0.099 & -5.585 & 0.013 \\
L 127-124        & 91.62  & 0.92 & N18  & 0.741  & 0.100 & -5.629 & 0.010 \\
UCAC4 195-119117 & 93.70  & 0.94 & N18  & 0.749  & 0.102 & -5.006 & 0.008 \\
GJ 836           & 94.25  & 0.94 & N18  & 0.752  & 0.102 & -6.195 & 0.019 \\
GJ 3382          & 95.00  & 0.95 & B22  & 0.755  & 0.103 & -5.698 & 0.019 \\
LP 728-71        & 99.66  & 1.00 & N18  & 0.774  & 0.109 & -5.632 & 0.011 \\
GJ 1057          & 102.00 & 1.00 & AD17 & 0.783  & 0.112 & -5.522 & 0.070 \\
GJ 667 C         & 103.10 & 0.60 & 1H   & 0.788  & 0.112 & -5.500 & 0.172 \\
GJ 628           & 108.70 & 1.50 & 1R   & 0.810  & 0.119 & -5.287 & 0.613 \\
GJ 3963          & 122.66 & 1.23 & N18  & 0.867  & 0.136 & -5.596 & 0.011 \\
GJ 447           & 124.95 & 1.90 & 1R   & 0.876  & 0.139 & -5.235 & 0.505 \\
PM J08446-4805   & 129.51 & 1.30 & N18  & 0.894  & 0.143 & -5.668 & 0.007 \\
L 210-11         & 136.92 & 1.37 & N18  & 0.924  & 0.153 & -5.998 & 0.018 \\
GJ 581           & 147.80 & 0.60 & 1A+R & 0.968  & 0.069 & -5.487 & 0.425 \\
GJ 699           & 149.50 & 0.60 & 1A+R & 0.968  & 0.069 & -5.490 & 0.634 \\
GJ 273           & 160.83 & 2.48 & 1R   & 0.968  & 0.069 & -5.273 & 0.520 \\
LP 855-60        & 154.89 & 1.55 & N18  & 0.997  & 0.174 & -5.963 & 0.025 \\
GJ 191           & 153.2  & 3.7  & 1S   & 1.061  & 0.046 & -5.609 & 0.495 \\
\enddata
\tablecomments{\footnotesize{$^1$ Measured in this study using data from: A = \textit{APT}, F = \citet{2020ApJS..246...11F}, H = \citet{2020AA...636A..74T} ,M = \textit{MEarth}, R = \textit{RCT}, S = \textit{Skynet}, Z = \textit{ZTF})  --  KS13: \citet{2013AcA....63...53K}  -- SM15: \citet{2015MNRAS.452.2745S}  -- SM16: \citet{2016AA...595A..12S}   -- N16: \citet{2016ApJ...821...93N}  -- SM17:  \citet{2017MNRAS.468.4772S}  --  AD17: \citet{2017AA...600A..13A}  -- N18:  \citet{2018AJ....156..217N}  -- SM18:  \citet{2018AA...612A..89S}  -- DA19: \citet{2019AA...621A.126D}  --  B22: \citet{2022ApJ...929...80B}
\newline ** Each star's $\log~\langle R_{\rm HK}\rangle$  value was determined using the combined data (depending on availability) from \citet{2013AA...551L...8P,2015MNRAS.452.2745S,2016AA...595A..12S,2017AA...600A..13A,2017MNRAS.468.4772S,2018AA...616A.108B,2018AA...612A..89S,2020AJ....160..269M,2021AA...652A.116P,2022ApJ...929...80B,2023AA...672A..37I,2023AA...671A.162M}; the Pleiades and Hyades values were obtained from \citet{2018MNRAS.476..908F}.
}}
\end{deluxetable*}
\end{longrotatetable}

\begin{longrotatetable}
\begin{deluxetable*}{lcccccccccc}
\tablecaption{Early M Dwarf Ly$\alpha$ Data \label{table:earlylya}}
\tablehead{\colhead{Star Name}  & \colhead{$P_{\rm{rot}}$ (days)}  & \colhead{err}  & \colhead{src}
    & \colhead{$\log$ Age (Gyr)}  & \colhead{err}  & \colhead{$\log~L_{\rm Ly\alpha}$}  & \colhead{err}
    & \colhead{$\log~(L_{\rm Ly\alpha} / L_{\rm bol})$}  & \colhead{err}}
\startdata
AU Mic    & 4.86  & 0.05 & K19  & -0.742 & 0.040 & -3.500 & 0.019 \\
HIP 23309 & 8.60  & 0.07 & M10  & -0.510 & 0.043 & -4.163 & 0.087 \\
GJ 410    & 14.52 & 0.02 & 1R   & -0.142 & 0.052 & -3.306 & 0.922 \\
GJ 49     & 19.45 & 0.06 & 1R   & 0.164  & 0.060 & -3.931 & 0.804 \\
GJ 649    & 21.42 & 0.20 & 1R   & 0.286  & 0.065 & -4.044 & 0.300 \\
GJ 3470   & 21.54 & 0.49 & Ko19 & 0.294  & 0.071 & -4.170 & 0.300 \\
GJ 887    & 33.00 & 0.91 & AD17 & 0.503  & 0.072 & -4.077 & 0.300 \\
GJ 205    & 33.72 & 0.13 & 1R   & 0.509  & 0.072 & -4.112 & 0.300 \\
\textbf{GJ 832}    & 38.10 & 0.11 & 1S   & 0.550  & 0.081 & -4.635 & 0.032 \\
GJ 176    & 39.30 & 0.10 & 1AR  & 0.561  & 0.083 & -4.515 & 0.026 \\
GJ 15 A   & 50.69 & 0.21 & 1R   & 0.668  & 0.107 & -4.302 & 0.300 \\
GJ 411    & 56.15 & 0.27 & D19  & 0.719  & 0.119 & -4.834 & 0.300 \\
\enddata
\tablecomments{\footnotesize{$^1$ Measured in this study using data from: A = \textit{APT}, M = \textit{MEarth}, R = \textit{RCT}, S = \textit{Skynet}, Z = \textit{ZTF}) -- K19: \citet{2019AA...622A..40K} -- Ko19: \citet{2019AJ....157...97K} -- DA19: \citet{2019AA...621A.126D}  -- AD17: \citet{2017MNRAS.464.3281W}  -- M10: \citet{2010AA...520A..15M}
}}
\end{deluxetable*}
\end{longrotatetable}

\begin{longrotatetable}
\begin{deluxetable*}{lcccccccccc}
\tablecaption{Mid-Late M Dwarf Ly$\alpha$ Data \label{table:midlatelya}}
\tablehead{\colhead{Star Name}  & \colhead{$P_{\rm{rot}}$ (days)}  & \colhead{err}  & \colhead{src}
    & \colhead{$\log$ Age (Gyr)}  & \colhead{err}  & \colhead{$\log~L_{\rm Ly\alpha}$}  & \colhead{err}
    & \colhead{$\log~(L_{\rm Ly\alpha} / L_{\rm bol})$}  & \colhead{err}}
\startdata
AD Leo (GJ 388)         & 2.24   & 0.02   & 1R, Mo08  & -0.764 & 0.019 & -3.576 & 0.013 \\
HIP 112312              & 2.355  & 0.005  & M10       & -0.758 & 0.019 & -3.170 & 0.075 \\
YZ CMi (GJ 285)         & 2.7758 & 0.0002 & Mo08      & -0.092 & 0.031 & -3.218 & 0.300 \\
GJ 729                  & 2.90   & 0.03   & KS12      & -0.727 & 0.019 & -3.483 & 0.300 \\
HIP 17695               & 3.87   & 0.01   & M10       & -0.673 & 0.019 & -3.356 & 0.064 \\
EV Lac (GJ 873)         & 4.3715 & 0.0002 & Mo08      & -0.052 & 0.031 & -3.769 & 0.095 \\
GJ 4334                 & 23.54  & 0.24   & N16, DA19 & 0.449  & 0.027 & -3.828 & 0.882 \\
LHS 2686                & 28.80  & 0.80   & N16, DA19 & 0.486  & 0.034 & -4.020 & 0.300 \\
GJ 674                  & 33.30  & 1.00   & KS07      & 0.505  & 0.037 & -4.060 & 0.300 \\
GJ 163                  & 56.04  & 2.47   & 1M        & 0.597  & 0.060 & -4.326 & 0.300 \\
GJ 588                  & 61.30  & 6.50   & AM15      & 0.618  & 0.070 & -4.153 & 0.300 \\
GJ 436                  & 71.40  & 0.40   & 1R        & 0.659  & 0.076 & -4.590 & 0.070 \\
GJ 849                  & 71.95  & 1.01   & 1R        & 0.661  & 0.077 & -4.432 & 0.300 \\
GJ 876                  & 87.30  & 5.70   & 1A        & 0.723  & 0.097 & -4.678 & 0.051 \\
Proxima Cen (GJ 551)    & 88.98  & 0.01   & 1S        & 0.730  & 0.096 & -3.795 & 0.008 \\
\textbf{GJ 667 C}                & 103.10 & 0.60   & 1H        & 0.788  & 0.112 & -4.227 & 0.083 \\
GJ 1214                 & 125.00 & 5.00   & Ma18      & 0.876  & 0.140 & -4.592 & 0.297 \\
GJ 1132                 & 129.15 & 3.87   & N18       & 0.893  & 0.144 & -4.535 & 0.249 \\
GJ 581                  & 132.50 & 6.30   & SM15      & 0.968  & 0.069 & -4.957 & 0.126 \\
Barnard's Star (GJ 699) & 149.50 & 0.60   & 1R        & 0.968  & 0.069 & -4.746 & 0.009 \\
Luyten's Star (GJ 273)  & 160.83 & 2.48   & 1R        & 0.968  & 0.069 & -4.213 & 0.300 \\
Kapteyn's Star (GJ 191) & 153.20 & 3.70   & 1S        & 1.061  & 0.046 & -4.969 & 0.300 \\
\enddata
\tablecomments{\footnotesize{$^1$ Measured in this study using data from: A = \textit{APT}, H = \textit{HARPS} - Trifonov, M = \textit{MEarth}, R = \textit{RCT}, S = \textit{Skynet}, Z = \textit{ZTF}) -- 
Mo08: \citet{2008MNRAS.390..567M}  --  
M10: \citet{2010AA...520A..15M} --
KS13: \citet{2013AcA....63...53K}  -- 
SM15: \citet{2015MNRAS.452.2745S}  -- 
N16: \citet{2016ApJ...821...93N}  -- Ma18: \citet{2018AA...614A..35M}  --  N18:  \citet{2018AJ....156..217N}  -- 
DA19: \citet{2019AA...621A.126D}
}}
\end{deluxetable*}
\end{longrotatetable}

\section{Software and third party data repository citations} \label{sec:cite}




%

\section{Acknowledgments}
The rotation periods (and therefore ages) of numerous stars in this
project were measured with the Robotically Controlled Telescope (RCT)
at Kitt Peak. I would like to thank the members of the RCT
Consortium for maintaining and operating the telescope.

This work was supported by Chandra grants AR8-19005X, and
GO0-21020X to Villanova University.
This research has made use of data obtained from the
Chandra Data Archive, and software provided by the Chandra
X-ray Center (CXC) in the CIAO application package. This
research has made use of data from XMM-Newton, an ESA
science mission with instruments and contributions directly
funded by ESA member states and NASA. We acknowledge the
use of public data from the Swift data archive. This research
has made use of data and/or software provided by the High
Energy Astrophysics Science Archive Research Center (HEASARC),
which is a service of the Astrophysics Science Division at
NASA/GSFC. 

\vspace{5mm}
\facilities{KPNO:RCT, CTIO:PROMPT, HST, Swift, CXO, XMM, ROSAT}


\software{astropy \citep{2013AA...558A..33A,2018AJ....156..123A,2022ApJ...935..167A},  
          Matplotlib \citep{Hunter:2007},
          Pandas \citep{mckinney-proc-scipy-2010},
          NumPy \citep{harris2020array},
          SciPy \citep{2020SciPy-NMeth}
          }





\bibliography{livredbib}{}
\bibliographystyle{aasjournal}



\end{document}